\documentclass[prl,reprint,twocolumn,superscriptaddress,floatfix,nofootinbib,longbibliography]{revtex4-1}
\usepackage{epsfig,amsmath,amssymb,color,comment,physics}
\usepackage[makeroom]{cancel}
\usepackage[caption=false]{subfig}
\usepackage{mathrsfs}
\usepackage{float}
\usepackage[countmax]{subfloat}
\usepackage[normalem]{ulem}
\usepackage[english]{babel}
\usepackage{dsfont}
\usepackage{braket}
\usepackage[bookmarks=true,colorlinks,linkcolor=OrangeRed,urlcolor=NavyBlue,citecolor=RoyalBlue]{hyperref}
\usepackage[dvipsnames]{xcolor}
\usepackage{graphicx}
\usepackage{cleveref}

\graphicspath{{./Figures/}}
\usepackage{amsmath, amssymb,latexsym,amsfonts}
\usepackage{orcidlink}
\usetikzlibrary{patterns}

\begin{document}
\title{Sign-problem landscape of dimer, loop, and ground state sectors of a U(1) quantum link model}

\author{Pallabi Dey${}^{\orcidlink{0009-0007-2460-4545}}$}
\email{pallabi.dey@saha.ac.in}
\affiliation{Theory Division, Saha Institute of Nuclear Physics, 1/AF Bidhan Nagar, 
Kolkata 700064, India}
\affiliation{Homi Bhabha National Institute, Training School Complex, Anushaktinagar, 
Mumbai 400094,India}

\author{Debasish Banerjee${}^{\orcidlink{0000-0003-0244-4337}}$}
\email{D.Banerjee@soton.ac.uk}
\affiliation{School of Physics and Astronomy, University of Southampton, University Road, SO17 1BJ, UK}

\author{Emilie Huffman${}^{\orcidlink{0000-0002-4417-338X}}$}
\email{ehuffman@wfu.edu}
\affiliation{Department of Physics and Center for Functional Materials, 
Wake Forest University, Winston-Salem, North Carolina 27109, USA}

\begin{abstract}
The fermion sign problem poses a formidable challenge to the use of Monte Carlo methods for
lattice gauge theories with dynamical fermionic matter fields. A meron cluster algorithm
recently formulated for gauge fields represented as spin-$\frac{1}{2}$ quantum links coupled 
to a single flavour of staggered fermions samples only two of the exponentially many Gauss law
(GL) sectors at low temperatures, allowing the simulation of those two GL sectors 
at zero temperature in polynomial time. In this article, we analytically identify GL sectors 
which can be simulated without encountering the fermion sign problem in arbitrary spatial 
dimensions. Using large-scale exact diagonalization and cluster Monte Carlo methods, we 
explore the nature of phases in the GL sectors dominating at zero temperature. 
The ground state lives in a superselection sector free of the sign problem, 
and maps to the quantum dimer model with mobile monomers. The usual zero-charge GL sector suffers 
from the fermion sign problem, and maps to the fully packed loop model with mobile monomers. 
The role of the magnetic energy in causing transitions between GL sectors is outlined. We expect 
our results to be valid for truncated Kogut-Susskind gauge theories, beyond quantum link models.
\end{abstract}

\date{\today}
\maketitle

\paragraph{Introduction.--} 
The fermion sign problem is one of the outstanding challenges in the ab-initio simulation
of quantum field theories (QFT) with fermionic matter using Markov Chain Monte Carlo (MCMC) 
methods \cite{Troyer2005}. Monte Carlo methods rely on the interpretation of the exponentiated 
Euclidean action as Boltzmann weights for importance sampling. With fermions involved, the 
Boltzmann weights in the occupation number basis of the fermions (or the field basis) are not 
explicitly positive definite for individual configurations, giving rise to the sign problem. 
Consequently, physics at finite baryon chemical potential \cite{Nagata2022} or the Hubbard model 
at finite doping \cite{Arovas2022} suffer from a severe fermion sign problem. 

There are multiple approaches to tackle sign problems 
\cite{Wu2005,Assaad2007,Chandrasekharan2010,Cristoforetti2012,Huffman2014,Huffman2016,Huffman2017,
Gattringer:2016kco,Wang2016,Li2019,Gantgen:2023byf}, but the meron concept 
\cite{Chandrasekharan1999,Liu2020} attacks the problem in a very direct way. The key idea is to 
first perform an analytic summation of certain fermionic worldline configurations to cancel those 
with equal and opposite signs, and subsequently to completely avoid generating such configurations 
during the importance sampling \cite{Banerjee:2021zed}. While the original meron proposal dealt 
with a single component of staggered fermions, generalizations involving multiple fermion species 
followed \cite{Chandrasekharan_2003}. Another very important direction involves incorporating 
gauge fields coupled to fermions \cite{Banerjee:2023cvs,PintoBarros:2024oph}. The microscopic 
Hamiltonian involving gauge fields has more constraints, due to superselection sectors specified by 
Gauss Laws, which make the application of the meron idea more challenging.  

\begin{figure}[!tbh]
\includegraphics[scale=0.6]{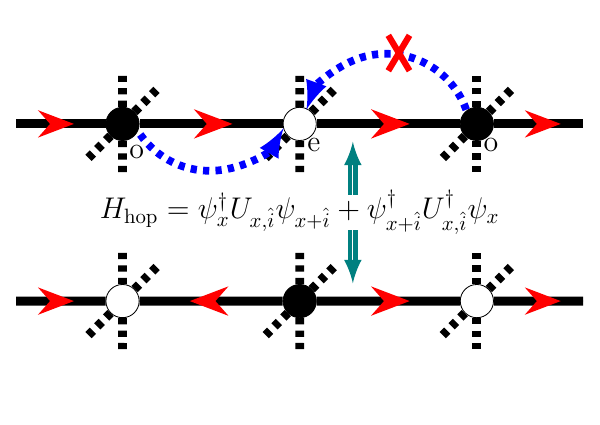}
\includegraphics[scale=0.2]{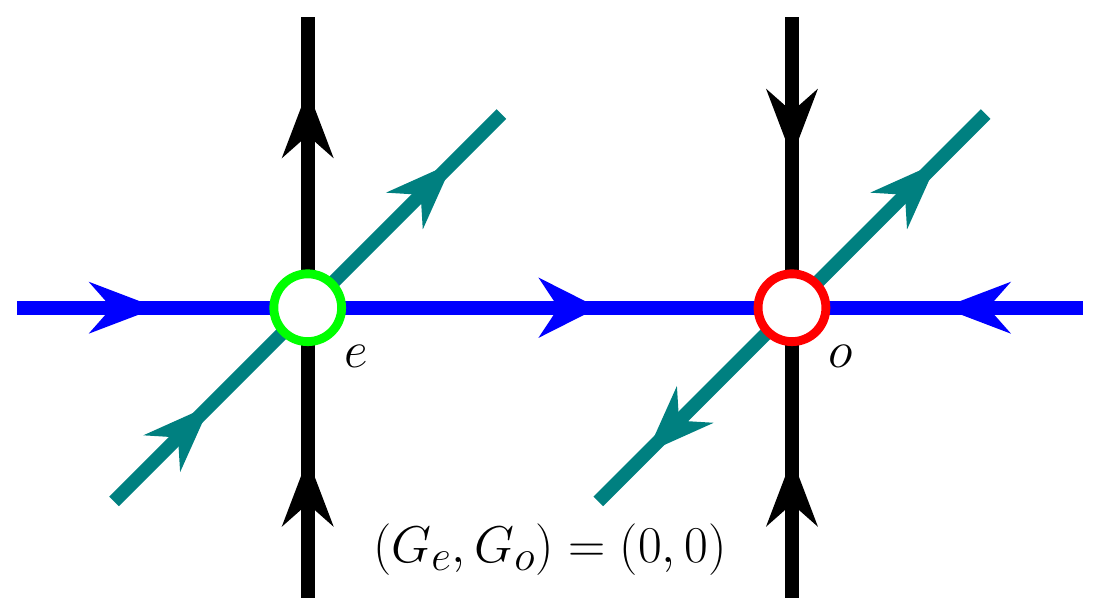} 
\hspace{0.5cm}
\includegraphics[scale=0.2]{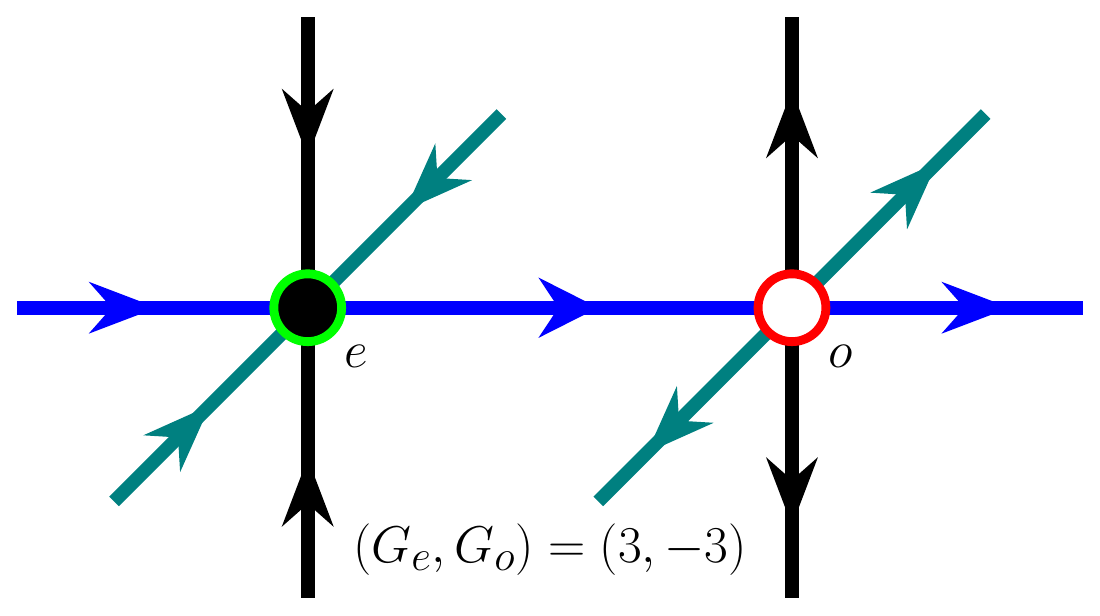}
\caption{(Top): The fermion hop from left to right is accompanied 
by a $\sigma^- (U^\dagger)$ operator on the link, causing the 
spin-$1/2$ electric flux to flip its orientation, while the right 
to left hop is accompanied by a $\sigma^+ (U)$ operator on the 
link and is thus constrained by the state of the flux on the link. 
(Bottom): Examples of electric flux configurations belonging to 
two different Gauss Law sectors. The direction of the arrows on 
the links indicates the flux to be $E_{x,\hat i} = \pm \frac{1}{2}$,
while the filled (empty) sites indicate the site to be occupied (empty).}
\label{fig:H-GL}
\end{figure}

 Adding gauge fields to fermions is a natural next step towards studying 
quantum chromodynamics. Other motivations 
include the rapidly evolving fields of analog and digital quantum simulator experiments 
which can simulate lattice gauge theories and constrained spin models in $d=1,2$ 
\cite{Martinez2016,Bernien2017,Ebadi2020,Yang2020,Zhou2021,Semeghini2021,Moss2023,Gonzalez-Cuadra2024,
Schuhmacher2025,Cobos2025,Luo2025,Davoudi2025,Klco2019,Maiti2024}
Some constrained models realized in analog quantum simulators are closely 
connected with quantum link models (QLMs) \cite{Surace2020}: e.g., the PXP model is the 
spin-$1/2$ quantum link Schwinger model. QLMs are generalized lattice gauge theories which 
realize exact gauge invariance with a finite-dimensional local Hilbert space for gauge 
links \cite{Chandrasekharan1997,Brower1999}, and thus can be efficiently encoded in quantum 
simulators and computers \cite{Wiese2021}. Moreover, by choosing an appropriate representation 
of the fermions or gauge fields, many QLMs can be mapped to pedagogical models in 
condensed matter physics, such as the quantum dimer model and the quantum spin ice 
\cite{Hermele:2004zz,Moessner2011,Banerjee2013,Banerjee2014}.
We explore two such exact mappings here: to the quantum dimer model 
and the fully packed loop model.

 Non-perturbative study of these models on large lattices requires advanced numerical techniques. 
While tensor network methods are extremely mature in $d=1$, their viability in $d>1$ spatial 
dimensions is still limited. Therefore, it is urgently required to expand the repertoire of 
classical Monte Carlo algorithms which can address the static properties of these theories. 
Such investigations are crucial not only for investigating ground state physics, serving as 
benchmarks for quantum computer implementations, but also influence planning possible investigations 
for real-time dynamics on quantum computers. Further, many of these models show novel phenomena 
beyond traditional lattice gauge theories \cite{Banerjee2013,Banerjee:2023pnb}, and could have 
intriguing technological implications.

Here we further explore $U(1)$-symmetric models of gauge fields and matter with meron cluster 
techniques and exact diagonalization (ED), specifically for $d=2$ and, $d=3$ spatial dimensions. 
The paper is organized as follows: after introducing the Hamiltonian and GLs, 
we analytically identify fermion-sign-problem free GL sectors by examining
fermion world lines, and we use a symmetry found in this analysis to establish a mapping between
QLM sectors with the quantum dimer and fully packed loop models. 
Using large-scale ED, the ground state physics in GL sectors that dominate at low temperatures is 
then studied. We also study transitions between the GL sectors as the 
couplings are varied. We conclude with a discussion on how our results are applicable to
truncated lattice gauge theories beyond QLMs.

\paragraph{QLM Hamiltonian and Gauss Law sectors.--}
 The original meron concept was applied to a single-component fermionic species with nearest
neighbour hopping, and interactions via a four-fermion coupling. When the gauge fields are coupled
to fermions, a fermion hop across a link to a neighbouring site changes the flux on the link. If
the theory uses a finite-dimensional flux representation, as in the QLM formulation, the hopping
is constrained by the flux on the link (see \cref{fig:H-GL} (top)). In the occupation number 
basis for the fermions and the electric flux basis for gauge fields, the Hamiltonian is: 
\begin{equation}\label{eq:H_U1}
\begin{aligned}
 H& = -t\sum_{x,i} [ (\psi^\dagger_x U_{x, \hat i} \psi_{x+\hat i} 
    + {\rm h.c.} ) 
    -  E_{x, \hat i}(n_{x+\hat i} - n_x ) ] \\
    &+ (V-t) \sum_{x,i}  \left(n_x - \frac{1}{2}\right)\left(n_{x+\hat i} 
    - \frac{1}{2}\right) - \frac{t \cdot d \cdot \cal{V}}{4}.
\end{aligned}
\end{equation}
 The first term causes the fermion hopping simultaneously with the gauge field $U_{x,\hat i} 
(U^\dagger_{x,\hat i})$ creating (annihilating) a unit of electric flux, $E_{x, \hat i}$, on the 
link joining sites $x$ and $x+\hat i$ as shown in \cref{fig:H-GL}. Next is the so-called 
\emph{designer} term equal in strength to the hopping and which simplifies the algorithm \cite{Banerjee:2023cvs}, but is not expected to change the
physics. A discussion of the model without the \emph{designer} term is given in SM 
Sec.~\hyperref[sec:without_designerterm]{VIII} \cite{SM}. The third term is a density-density 
interaction at neighbouring sites, equivalent to a four-fermi coupling. We imagine the 
system in $d$-dimensions with spatial volume $\cal{V}$.

\begin{figure}[!tbh]
\includegraphics[scale=0.4]{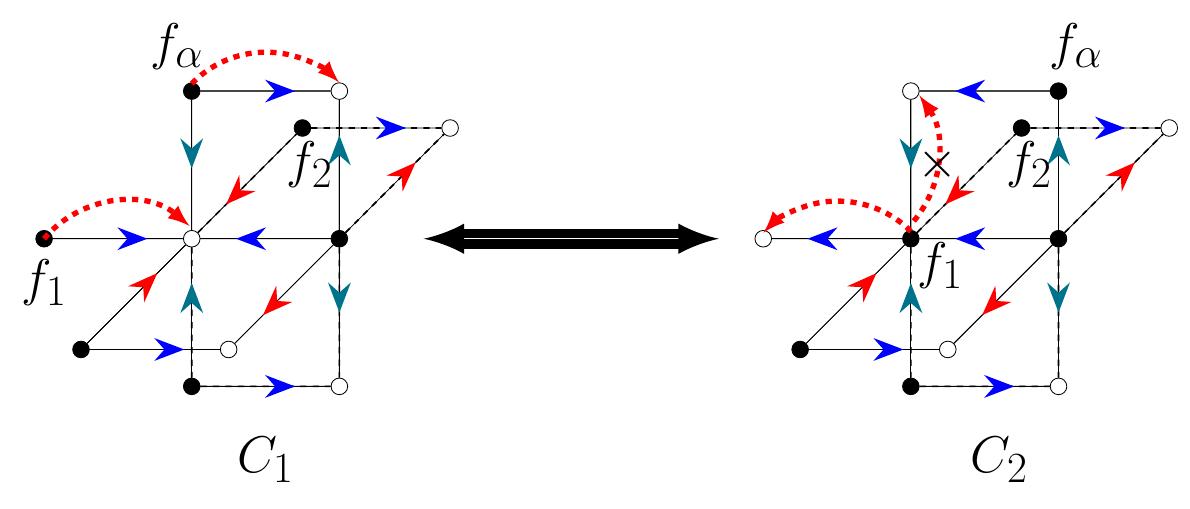}
\caption{The orientation of the gauge links in the sector $(3,-3)$ in $d=3$ do not allow the 
movement of fermions beyond one lattice spacing. Thus, positions of $f_1$ and $f_2$ cannot be 
switched by the action of the Hamiltonian.}
\label{fig:GL3d3}
\end{figure}

   Here we use the spin-$\frac{1}{2}$ representation for gauge fields, with two states for
each link. Although any spin-$S$ representation maintains the $U(1)$ gauge 
symmetry, the only known meron-cluster algorithm, for $d>1$, is for spin-$\frac{1}{2}$.
This is also a maximally constrained system, likely to display exotic physics. 
In $d=1$, the continuum limit can be obtained with relatively low spin representations
\cite{PhysRevD.106.L091502}. The local $U(1)$ symmetry transformation is generated by the 
following Gauss Law operator:
\begin{equation} \label{eq:GL}
    G_x = n_x + \left( \frac{(-1)^x - 1}{2} \right) - \sum_{i} (E_{x,\hat i} - E_{x-\hat i,\hat i}), 
\end{equation}
 where $(-1)^x$ is the site parity, 
 and the second term addresses the staggered fermion occupation. 
 Since $[G_x, H]=0$, $G_x$ splits the physical Hilbert space into superselection sectors, 
labelled by its eigenvalues. The staggered fermion occupation in \cref{eq:GL} is necessary 
for preserving the global charge conjugation symmetry and a charge neutral vacuum. Consequently, 
we use $(G_e,G_o)$ to denote even and odd-site GL sectors locally. In \cref{fig:H-GL} (bottom), 
we show examples of flux configurations in two different GL sectors, $(0,0)$ and $(3,-3)$, which 
play a role in our investigation in $d=3$.

\begin{figure*}[t]
  \centering
  \begin{minipage}[t]{0.60\textwidth}
    \vspace{0pt}
    \centering
    \includegraphics[width=\linewidth,height=0.30\textheight,keepaspectratio=false]{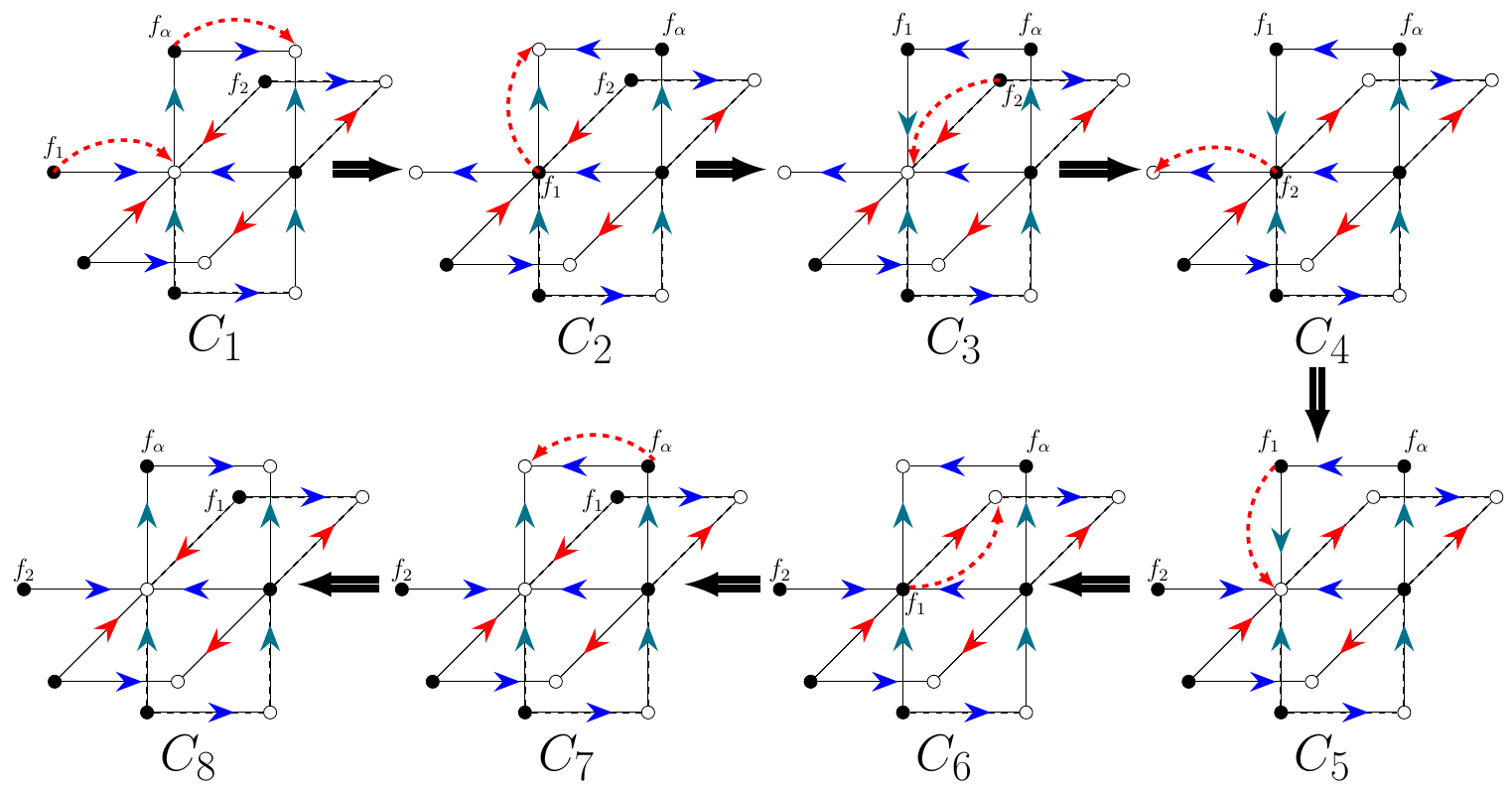}
  \end{minipage}%
  \hfill
  \begin{minipage}[t]{0.36\textwidth}
    \vspace{0pt}
    \centering
    \includegraphics[width=\linewidth]{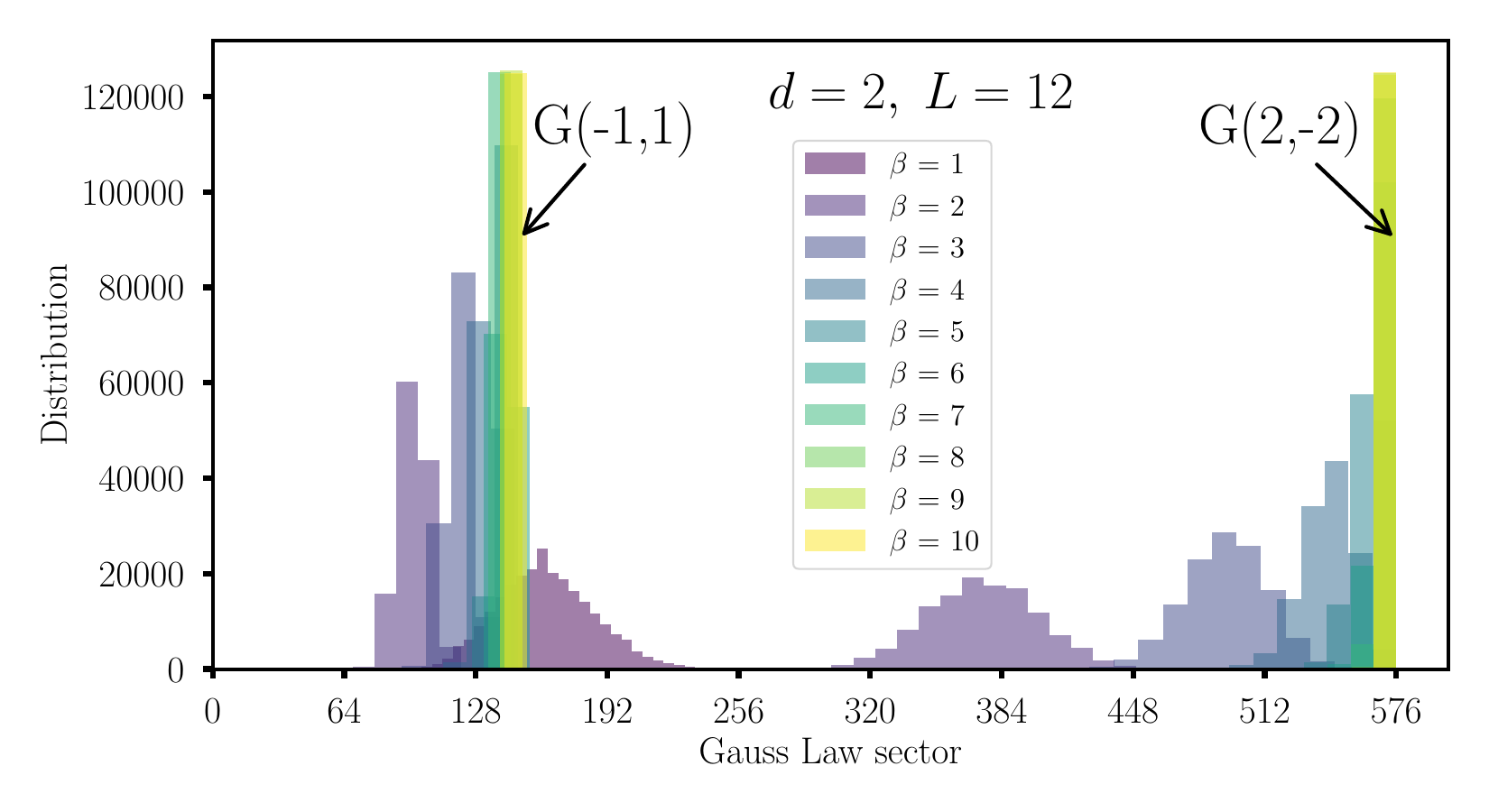}\\
    \includegraphics[width=\linewidth]{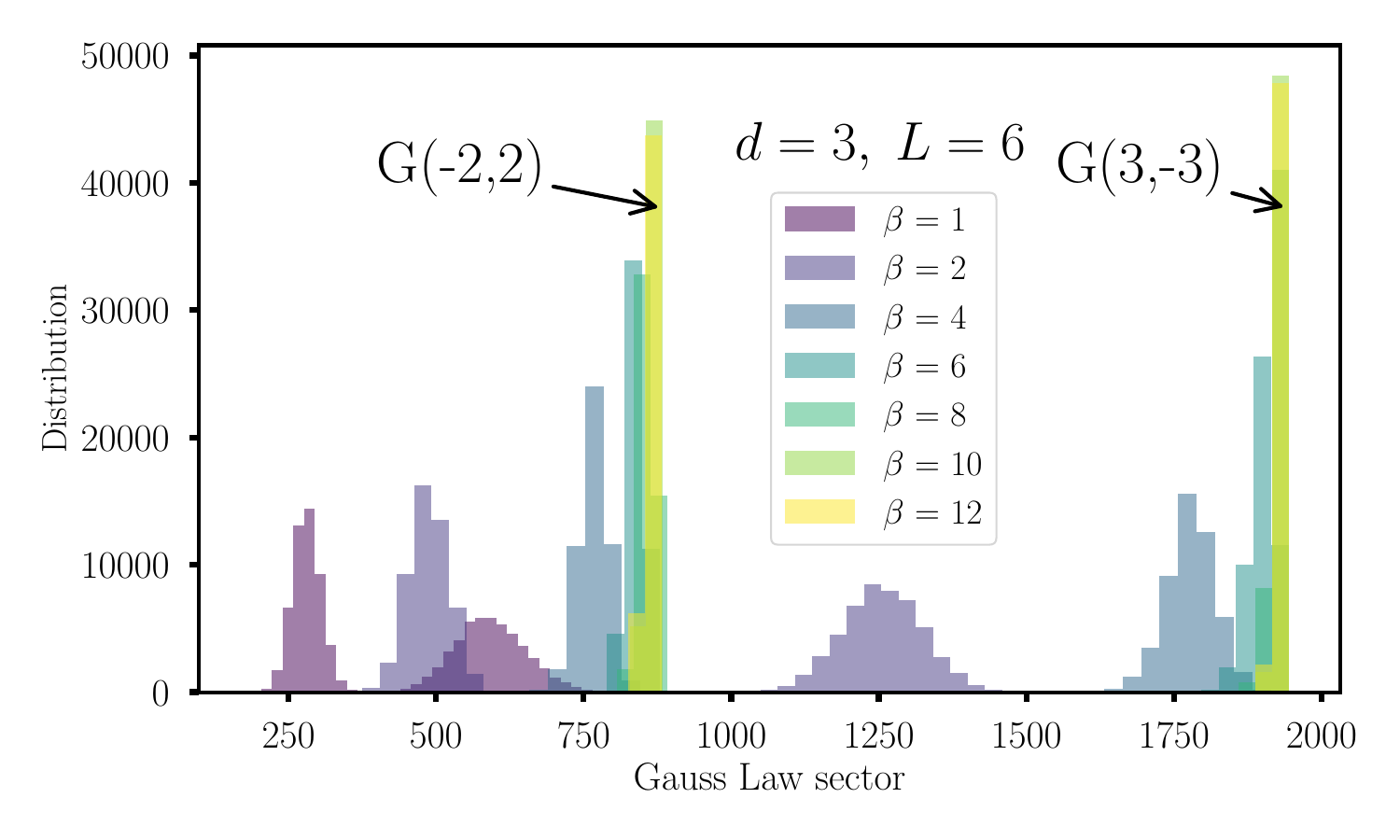}
  \end{minipage}
  \caption{(Left): The GL constraints in the sector $(2,-2)$ in $d=3$ are relaxed enough
  to allow fermions $f_1$ and $f_2$ to exchange positions with each other following the 
  general prescription described in the text. (Right) The different GL sectors that are
  sampled by the QMC algorithm at different $\beta$. For low temperature (large $\beta$)
  only the GL $(d,-d)$ and its shifted partner arises.}
  \label{fig:GL2d3}
\end{figure*}

 Postponing a discussion of global symmetries to the SM Sec.~\hyperref[sec:symm]{I} \cite{SM}, 
we instead concentrate on the consequence of the discrete chiral symmetry, where all 
fermion and gauge operators are translated by one lattice spacing in a given direction. 
The two-site GL $(G_e, G_o) = (n - \nabla \cdot E, n - 1 - \nabla \cdot E)$, transforms 
as $G_e \rightarrow G_o + 1, G_o \rightarrow G_e - 1$. The Hilbert space for $d=2$ 
(square lattice) in the GL sector $(0,0)$ is constructed by solving $n - \nabla \cdot E = 0$ 
and $n - \nabla \cdot E = 1$ for even and odd sites respectively. The first equation 
has 6 (4) solutions with $n=0 \;(n=1)$, while the second gives 4 (6) solutions with 
$n=0\;(n=1)$, and thus we have $10\;(10)$ solutions for the GL sector $(0,0)$ on even 
(odd) sites when counting independently. For the shifted GL sector $(1,-1)$, the equations 
for the even and odd sites are interchanged, but the number of solutions are the same. 
This, together with the fact that the Hamiltonian does not connect the $(0,0)$ and the 
$(1,-1)$ sector, establishes an isospectral relation between the two sectors. Similarly, 
the pairs $\left[ (2,-2)~{\rm and}~(-1,1)\right]$ and $\left[(3,-3)~{\rm and}~(-2,2) \right]$ 
are isospectral, independent of spatial dimensions. More details are given in SM 
Sec.~\hyperref[sec:GLsoln]{II} \cite{SM}.

 An important connection to these Gauss law sectors is an exact mapping between
quantum dimer model (QDM) \cite{Yan:2019vye} and the $(-1,1)$ sector, and an exact mapping between 
the fully packed quantum loop model (FPL) and the $(0,0)$ sector \cite{Ran2023} 
\emph{--both with dynamical monomers}. The QDM and the FPL models without monomers map 
to $G_x=(-1)^x$ and $G_x=0$ sectors of pure gauge theories, respectively \cite{Banerjee2013}. Mobile 
monomers are related to dynamical matter fields, and our gauge theory formulation identifies the 
relevant superselection of the Hilbert space, where the theories reside, as explained in SM 
Sec.~\hyperref[sec:mappings]{IX} \cite{SM}.

\paragraph{Sign Problem in different sectors.--} 
The spin-$1/2$ gauge links and the GL prevent any permutation of fermionic 
worldlines in $d=1$, leading to no sign problem in any GL sector. This restriction is absent 
in $d>1$, so we explore the nature of the sign problem in $d=2,3$ in different GL sectors. 
The surprising analytic results are supplemented with numerical checks in SM Secs.
\hyperref[sec:meron]{V},~\hyperref[sec:data]{VI} \cite{SM}.

Our key result shows \emph{the absence of the sign problem in the Gauss Law sector 
$(d,-d)$ and its shift partner $(-d+1,d-1)$}. Stated differently, the partition functions in 
these sectors are identical if the fermions are replaced by (hardcore) bosons. The analytic 
proof proceeds by identifying two fermions in a spatial configuration and interchanging their 
positions by acting with the Hamiltonian, keeping the positions of all other fermions fixed. 

\begin{figure}[!tbh]
\includegraphics[scale=0.2]{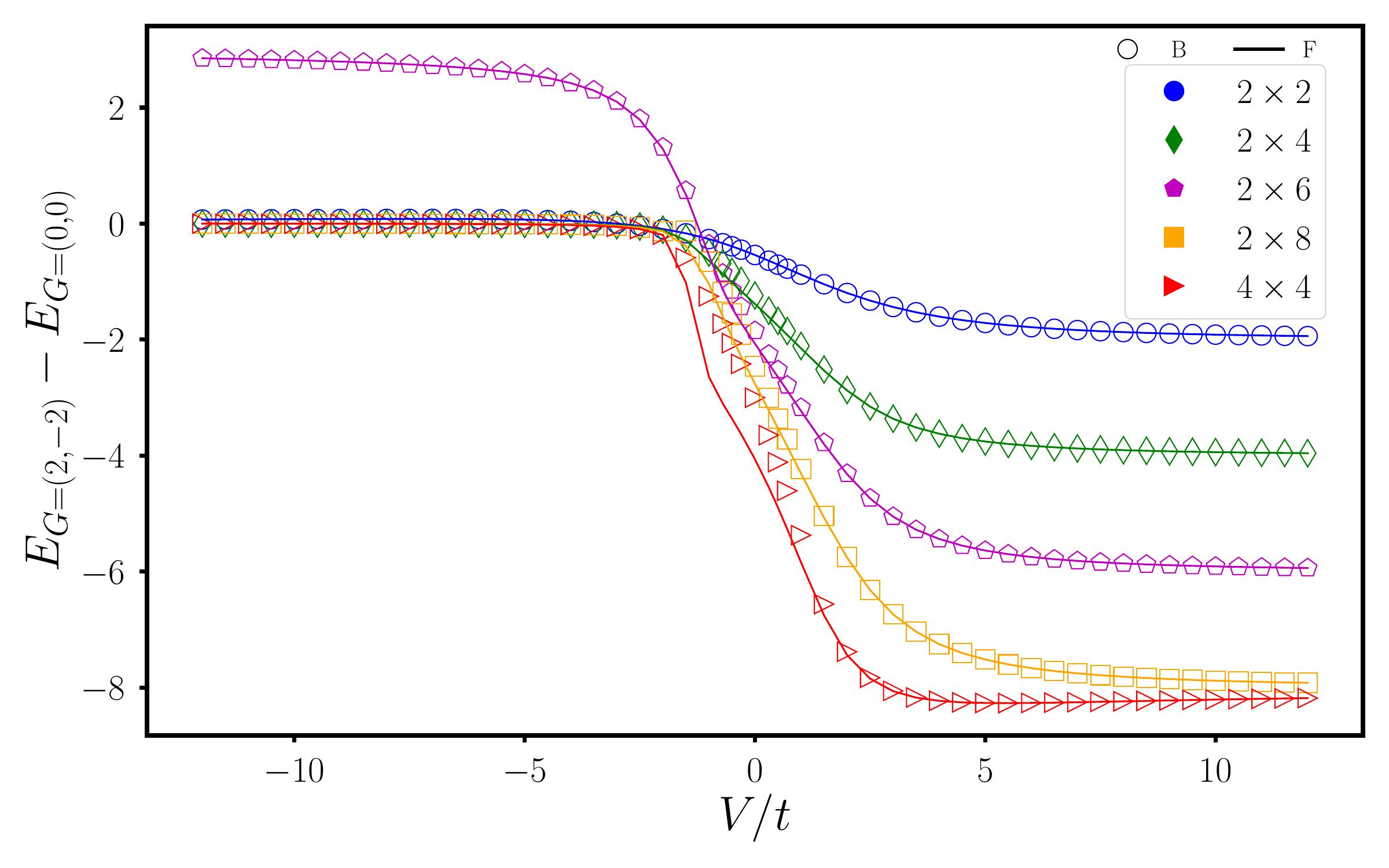}
\caption{Ground state energy difference between the GL sectors $(2,-2)$ and $(0,0)$
for bosons (open symbols) and fermions (solid lines) respectively in $d=2$. While the fermionic 
and bosonic results are the same for the sector $(2,-2)$, there is a difference in the 
$(0,0)$ sector for $V/t \sim 0$, leading to the deviation between the two sets of data 
in that region.}
\label{fig:delE}
\end{figure}

 Consider the sector $(3,-3)$, which has $(7,7)$ allowed solutions to the GL constraints, 
one of which is shown in \cref{fig:GL3d3}. We identify $(f_1,f_2)$ as two fermions with 
positions $(r_1,r_2)$, and we also identify an auxiliary fermion $f_\alpha$ at position 
$r_\alpha$ which we will use to facilitate the interchange in positions of $(f_1,f_2)$.
The procedure is to move $f_1$ to $r_\alpha$ (by shifting $f_\alpha$), while $f_2$ is moved to 
$r_1$. It is then completed by pushing $f_1$ to $r_2$, and $f_\alpha$ is restored to its original 
location. As shown in \cref{fig:GL3d3}, just after a single hop of $f_1$, we encounter links 
(fixed by GL) which forbid further hopping of $f_1$ --- only the reverse hop back to $r_1$ 
is allowed. We have verified this to be the case for \emph{all} configurations locally in 
the GL sector $(3,-3)$ (and by extension to its shift partner). This is in contrast to the 
case of the sector $(2,-2)$ shown in \cref{fig:GL2d3} (left), where the above procedure 
works out. Thus, configurations $C_1$ and $C_8$ differ by an overall sign when computations 
are performed in this basis. It results in a severe sign problem in a QMC sampling, which 
can be potentially solved with the meron concept.

 In the SM Sec.~\hyperref[sec:GL0]{III} \cite{SM}, this argument is outlined for the $(0,0)$ sector, 
particularly relevant in particle physics. Thus, in $d=2$: the sector $(2,-2)$ has no sign 
problem, but the $(0,0)$ has one (and is identical to the $(1,-1)$ sector). In $d=1$ this 
explains the absence of merons in \cite{Banerjee:2023cvs}. 
Fig \ref{fig:GL2d3} (right) shows distributions of the different GL sectors obtained in the 
QMC algorithm at different temperatures for both $d=2,3$ \cite{zenodo_data}. Details of the QMC algorithm used
are provided in the SM Secs.~\hyperref[sec:qmc]{IV},~\hyperref[sec:meron]{V},~\hyperref[sec:data]{VI} \cite{SM}.

\begin{figure}[!tbh]
\includegraphics[scale=0.198]{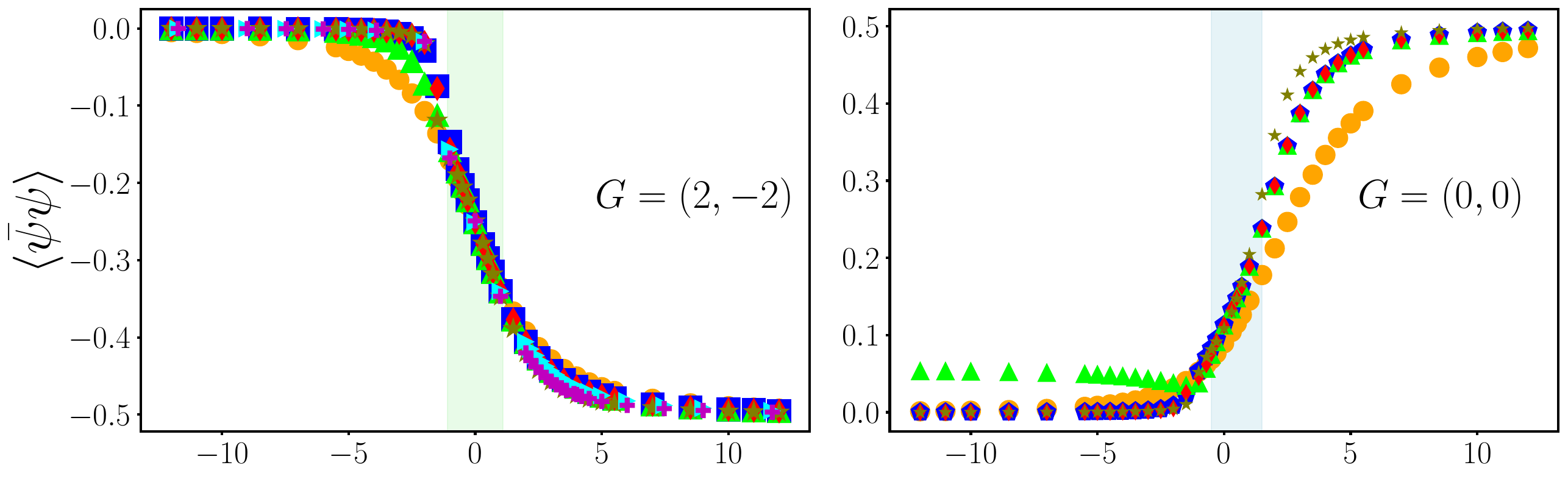}
\includegraphics[scale=0.24]{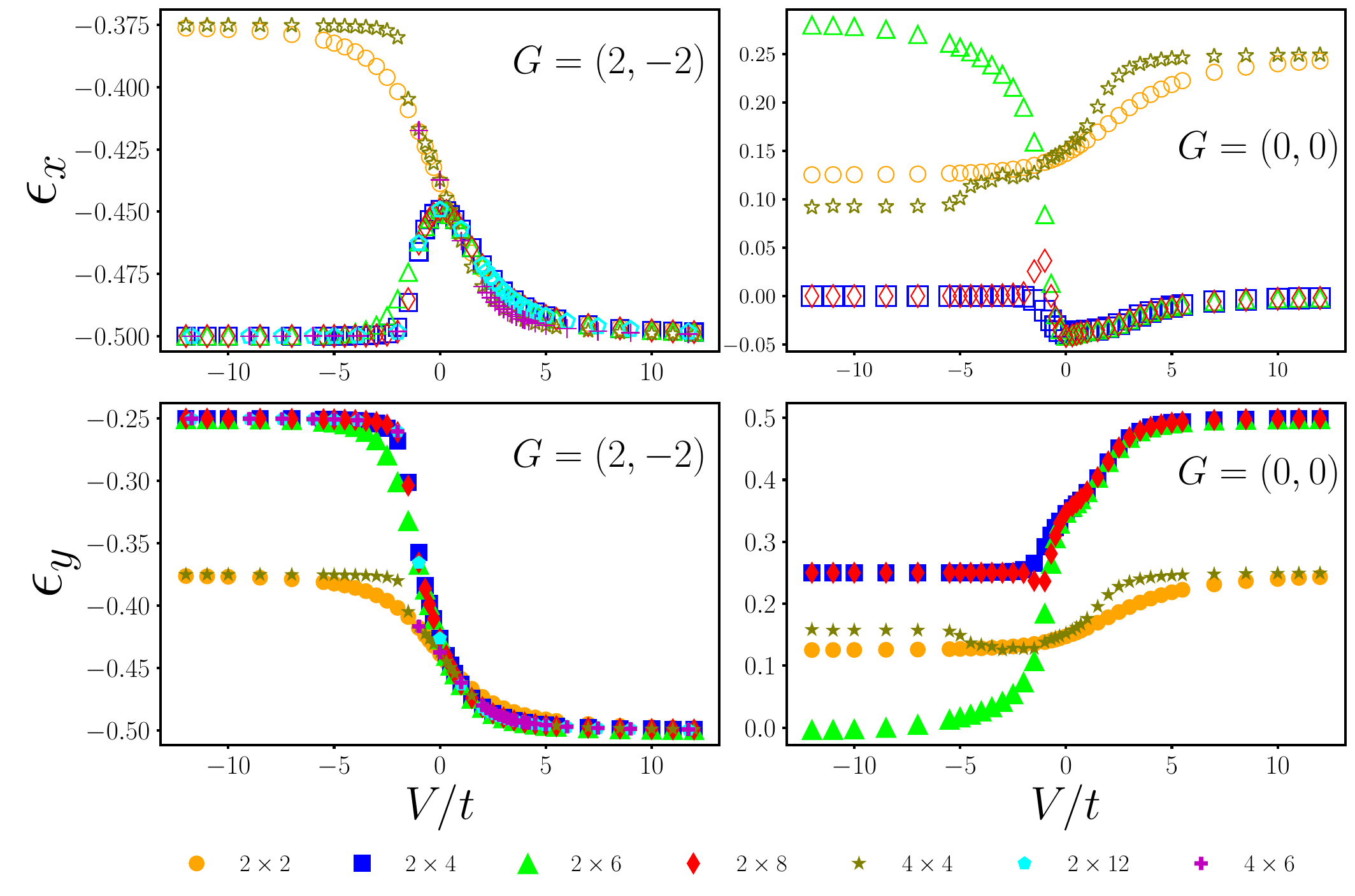}
\caption{ Top two figures show $\braket{\bar{\psi} \psi}$ vs $V/t$ in the two GL sectors
in $d=2$. The $\mathbb{Z}_2$ chiral symmetry breaks for $V/t \gg 0$ in both sectors, while
phase separation occurs for $V/t \ll 0$. The shaded band for $V/t \sim 0$ indicates a region 
where a liquid phase is suspected. The bottom two panels show $\epsilon_{x,y}$ for each GL 
sector in $d=2$. In $(2,-2)$ the symmetric ordering of the flux causes both $\epsilon_{x,y}$ to
reach the same value for $V/t >0 $ while due to phase separation at $V/t < 0$, $\epsilon_x$ and 
$\epsilon_y$ have different values, also depending on whether it is a ladder or a square geometry.
In GL sector $(0,0)$, the ordering of the flux requires a larger unit cell, and is thus sensitive 
to a multiple of 2 vs 4.}
\label{fig:GLPhys}
\end{figure}

\paragraph{Low energy physics.--} 
The next relevant question is the role of each GL sector at $T \approx 0$. 
 In $d=2$, using large-scale ED, we study the ground state at different $V/t$ regimes for both 
(hardcore) bosonic and fermionic matter up to 48 DOF (16 matter sites and 32 gauge links).  
A Krylov-space based ED gives acess to system sizes till 72 DOF (i.e., lattices $12 \times 2$, 
$6 \times 4$). To identify the GL sector of the ground state, \cref{fig:delE} plots the difference 
in ground state energies of the two GL sectors indicated in the subscript, 
$E^0_{(2,-2)}-E^0_{(0,0)}$, for bosonic and fermionic matter (the other sectors are higher in energy).
The difference arises only in the regime $V/t \sim 0$, when particles can maximally hop making their
exchange statistics manifest. Large (whether positive or negative) $V/t$ values fix $n_x$ values, 
making the bosonic or fermionic nature irrelevant. Except for a finite size effect on the $2\times6$ 
ladder, $E^0_{(2,-2)}$ is lower in energy for almost the entire regime, except for negative $V/t$ 
when it is degenerate. 

\begin{figure}[!tbh]
\includegraphics[scale=1.0]{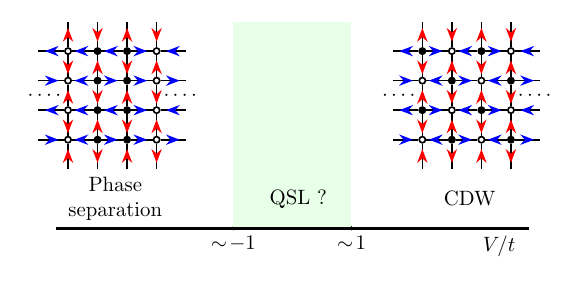}
\caption{Sketch of the phase diagram in the Gauss Law sector 
$(2,-2)$ as a function of the coupling $V/t$ in $d=2$. The existence of a quantum 
spin liquid (QSL) phase in between two symmetry breaking phases has been motivated 
using a finite size scaling provided in the SM Sec.~\hyperref[sec:FSS]{VII} \cite{SM}.}
\label{fig:phaseDiag}
\end{figure}

The ground state values of the chiral condensate ($\braket{\psi^\dagger \psi}$) and 
the sum of staggered electric flux, $\epsilon_{x,y}$, defined in \cref{eq:observable}, 
are plotted in \cref{fig:GLPhys}. Pure fermions at $V/t \gg 0$ would give rise to 
the charge-density wave (CDW) phase (spontaneous breaking of the $\mathbb{Z}_2$ chiral 
symmetry). Gauge fields explicitly break the $\mathbb{Z}_2$ symmetry and the two CDW 
vacua belong to two different GL sectors, with equal and opposite (non-zero) values 
of the chiral condensate. For $V/t \ll 0$, both sectors have phase separated ground 
states featuring neighbouring sites that are empty on side of the lattice, and filled 
on the other, causing the condensate to vanish. There is a narrow intermediate regime 
characterized by matter hopping, and finite size scaling (see SM Sec.~\hyperref[sec:FSS]{VII} \cite{SM}) 
indicates that it may be a liquid phase. For the GL sector $(0,0)$ similar results 
have been obtained in \cite{Cardarelli:2019toq,Hashizume:2021qbb,NSSrivatsa:2025jhh}. 

The electric fluxes (see \cref{fig:GLPhys} (bottom)) corroborate this understanding. 
The $\mathbb{Z}_2$ symmetry breaking for $V/t \gg 0$ causes both the staggered fluxes 
in $x$ and $y$ asymptote to the same value (in the $(0,0)$ sector the flux ordering 
is sensitive to $L_y$ being a multiple of 2 or 4). For $V/t \ll 0$, the phase separation 
makes $\braket{\epsilon_x}$ and $\braket{\epsilon_y}$ depend on whether the system 
is square or rectangular, indicating a spontaneous breaking of discrete rotation 
symmetry. More details are given in the SM Sec.~\hyperref[sec:FSS]{VII} \cite{SM}, while
\cref{fig:phaseDiag} shows a schematic phase diagram.

\begin{figure}[!tbh]
\includegraphics[scale=0.3]{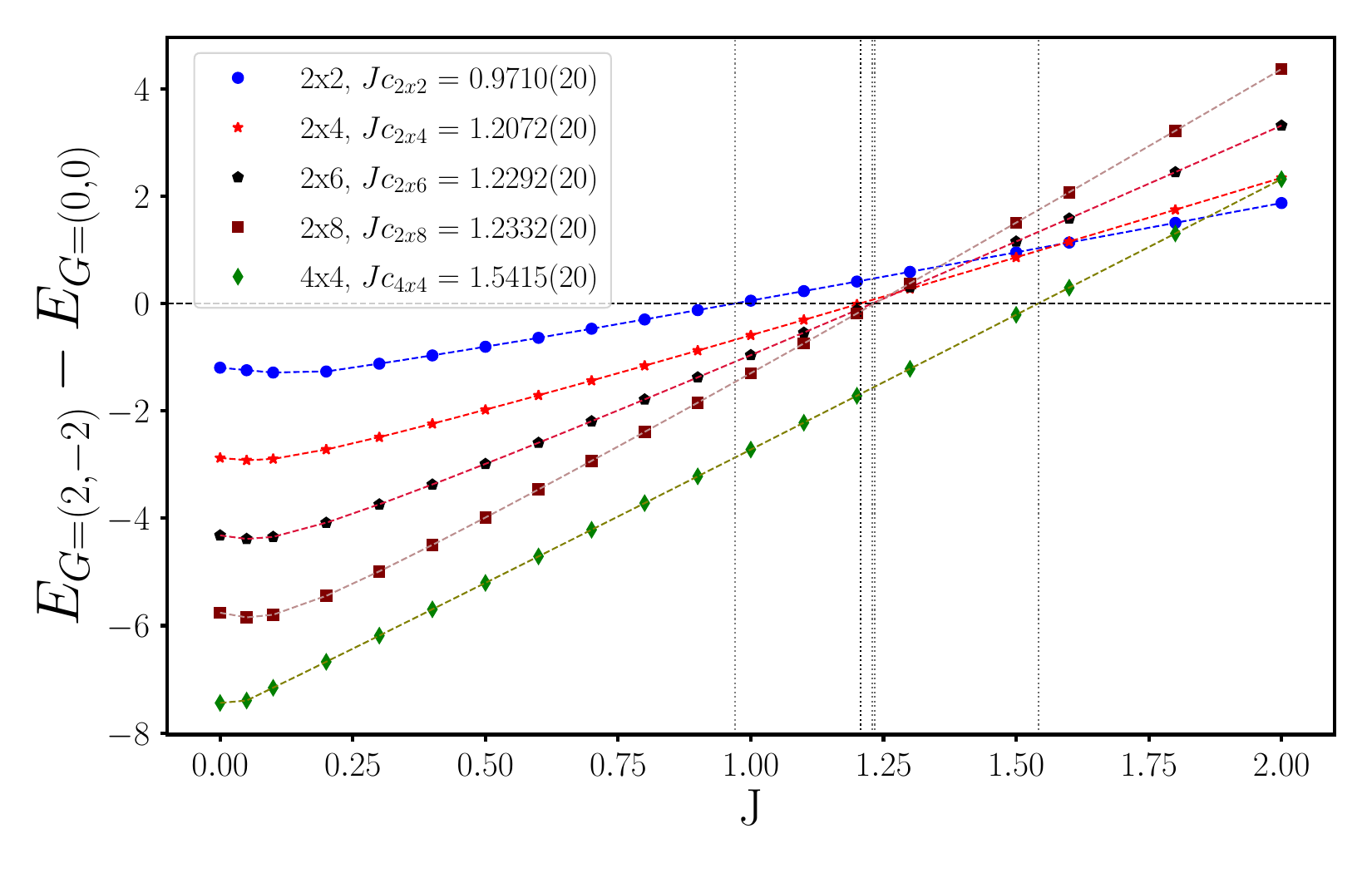}
\caption{Transition between the $(2,-2)$ and $(0,0)$ GL sectors as the magnetic 
coupling is increased.}
\label{fig:plaqEffect}
\end{figure}

 We have also examined the inclusion of the magnetic energy term,
$-J \sum_\square (U_\square + U^\dagger_\square)$, to the Hamiltonian in \cref{eq:H_U1},
for fermionic matter at $V=2t$. Fig \ref{fig:plaqEffect} (analogous to 
\cref{fig:delE} for both ladder and square systems) shows that although the GL sector $(2,-2)$ 
is the ground state across \emph{all} GL sectors
 when $J$ is small, for $J\geq J_c$ where the magnetic field dominates, the
sector $(0,0)$ becomes lower in energy and the system undergoes a phase transition.
For ladder systems, we estimate $J_c \approx 1.23(1)$, shown as a dashed vertical line 
in \cref{fig:plaqEffect}. For square systems, $J_c$ is larger, but we expect it to still be 
$O(1)$ based on the scaling for $2\times2$ and $4\times 4$ systems.
The $6\times6$ lattice has a total of 108 DOF, and is beyond the scope of our ED. The meron
algorithm does not yet work for $J \neq 0$.

\paragraph{Conclusions and Outlook.--} In this article, we have analyzed the fermion sign 
problem in different GL sectors of Abelian gauge theories with a finite-dimensional Hilbert
space. Although the explicit examples considered were spin-$\frac{1}{2}$ quantum links, the
methods developed in the paper can be straightforwardly applied to a spin-$S$ link model
or a truncated Kogut-Susskind Hamiltonian. Our results are relevant for quantum simulator 
experiments, particularly using Rydberg atoms \cite{Gonzalez-Cuadra2024,Osborne2025,
Bharadwaj2025,Crippa2024}, and perhaps design quench experiments to understand dynamical 
properties of this model. 
Our model in the sign-problem-free $(-1,1)$ sector maps to the monomer-dimer 
model experimentally realised in \cite{Karch:2026whf}, relevant for studies of quantum spin liquids.

Constructing the meron algorithm in GL sectors $(0,0)$, which maps 
to the FPL model with monomers, stands as the next challenge. Currently this sector has a 
sign problem and can only be reliably studied with quantum simulators, making this an area 
of study with potential for quantum advantage even in the near term. Extending the algorithm 
to multiple fermion flavours can help us verify existing results in the sign-problem-free 
extensions of quantum electrodynamics (QED3) in $d=2$ in
\cite{Bashir:2008fk,Assaad:2016flj,Chester:2016ref,Pufu:2013eda,Boyack:2018zfx,Xu2019}.

\paragraph{Acknowledgements.--} We would like to thank Monika Aidelsburger, Joao Pinto Barros, 
Thea Budde, Shailesh Chandrasekharan, Arnab Sen, and Uwe-Jens Wiese for useful discussions 
D.B. would like to thank STFC (UK) consolidated grant ST/X000583/1 and continued support 
from the Alexander von Humboldt Foundation (Germany) in the context of the research fellowship 
for experienced researchers. We thank computing resources of SINP and DiRAC (computational
facility for STFC) essential to derive these results.

\emph{Data availability.--} Data for this Letter are available on Zenodo~\cite{zenodo_data}.

\bibliography{ref}
\newpage
\appendix 

\section{Supplementary Material}
\subsection{I. Global and Gauge Symmetries} \label{sec:symm}
 We discuss the global symmetries of the Hamiltonian (see also \cref{eq:H_U1}):
\begin{equation}
\begin{aligned}
 H &= -t\sum_{x,i} \left[ \left(\psi^\dagger_x U_{x, \hat i} \psi_{x+\hat i} 
    + {\rm h.c.} \right) 
    -  E_{x, \hat i}\left(n_{x+\hat i} - n_x \right) \right] \\
    &+ (V-t) \sum_{x,i}  \left(n_x - \frac{1}{2}\right)\left(n_{x+\hat i} 
    - \frac{1}{2}\right) - \frac{t \cdot d \cdot \cal{V}}{4}.
\end{aligned}
\end{equation}
which is split into different superselection sectors labelled via local
charges labelled by the Gauss Law (see also \cref{eq:GL}):
\begin{equation} 
    G_x = n_x + \left( \frac{(-1)^x - 1}{2} \right) - \sum_{i} (E_{x,\hat i} - E_{x-\hat i,\hat i}), 
\end{equation}

 Note that this differs from the usual staggered fermion formulation by
omitting the staggered phase $\eta_{x,i} = (-1)^{\sum_{j < i} x_j}$ in the
kinetic energy term for the hopping fermions; where we have used the notation
$x = (x_1, x_2, \cdots, x_d)$. While the $\eta$ phase is included to ensure relativistic
dispersion for fermions in pure fermionic theories, this seems to be not essential when gauge
fields are coupled to the fermions \cite{Creutz:2013ofa,Xu2019}.

 The major discrete symmetry that plays a key role in our analysis is the shift 
symmetry, $\mathcal{S}_k$, which is a discrete translation of all fields by one 
lattice unit along the $k$-th spatial direction. This is the discrete chiral symmetry
for this model. The symmetry is implemented on the operators as:
\begin{equation}\label{eq:shiftsym}
    \begin{aligned}
       {}^{\mathcal{S}^{k}}\!\psi_x &= \psi_{x+\hat k}, 
        {}^{\mathcal{S}^{k}}\!{{U_{x,\hat i}}} = U_{x+\hat k, \hat i}, 
        {}^{\mathcal{S}^{k}}\!{{E_{x,\hat i}}}  = E_{x+\hat k, \hat i}.        
    \end{aligned}
\end{equation}
However, under the action of \cref{eq:shiftsym}, the Gauss law operator at site 
$x$ transforms as, 
\begin{equation}
    \begin{aligned}
        G_x \xrightarrow{\mathcal{S}^k} G_{x+\hat{k}} + (-1)^x.
    \end{aligned}
\end{equation}
This necessitates the need of the two-site labelling used in the paper. Under this 
transformation, each Fock state in the Gauss law sector $G(e, o)=(0,0),(2,-2),(3,-3)$ 
is mapped to a unique Fock states in a corresponding GL sector, $(1,-1),(-1,1),(-2,2)$ 
respectively. These sectors related by the shift symmetry $\mathcal{S}^k$ and are 
therefore physically equivalent since the two sectors are isomorphic. Further, in 
$d>1$, shifts by single lattice spacings in two \emph{different} directions generate
a discrete $\mathbb{Z}_2$ flavour symmetry. 

 Shifts by two lattice spacings in the same direction generate the usual 
\emph{translation} symmetry, $T_k$, in the $k$-th direction. In addition, we have
usual discrete rotational symmetries, either by a $\pi$-rotation or a $\pi/2$
rotation depending on whether the system is rectangular or a square.

 Finally, charge conjugation is an internal symmetry: for staggered fermions, this
is defined acting in a particular direction, as follows:
\begin{equation}\label{eq:chgconjsym}
    \begin{aligned}
       {}^{\mathcal{C}^{k}}\!\psi_x &= \psi^\dagger_{x+\hat k}, 
        {}^{\mathcal{C}^{k}}\!{{U_{x,\hat i}}} = U^\dagger_{x+\hat k, \hat i} , 
        {}^{\mathcal{C}^{k}}\!{{E_{x,\hat i}}}  = -E_{x+\hat k, \hat i}.        
    \end{aligned}
\end{equation}

The different phases realized in the regimes studied break either the shift symmetry
for large values of $V/t$ or the point group symmetry (the translation and rotational
symmetries) at large negative values of $V/t$, as discussed in the main text. Finite
size scaling of energy gaps as a diagnostic for these phases are discussed in the last
section of this appendix~\hyperref[sec:FSS]{VII}.

\subsection{ II. Solutions in different GL sectors} 
\label{sec:GLsoln}
For the square lattice, the local constraint $n - \nabla \cdot E = g$ restricts the 
allowed electric field values on the $2d$ links around a site, with the admissible 
configurations determined by $n - g$ and by whether the site is even or odd. For the 
GL sector $(2,-2)$, the local Hilbert space is constructed by solving $n-\nabla \cdot E=2$ 
on even sites and $n-1 -\nabla \cdot E=-2$  on odd sites. The first equation gives 1 (4) 
allowed solutions with $n=0 \ (n = 1)$ on even sites, while the second equation yields 4 (1) 
solutions for $n=0 \ (n = 1)$ on odd sites, resulting in $(5,\ 5)$ allowed solutions on 
(even, odd) sites for $d=2$. In the shifted GL sector $(-1,1)$, the Gauss law equations 
on even and odd sites are swapped relative to the sector $(2,-2)$, that leads to the same 
number of solutions.

In case of the cubic lattice $d=3$, each site is connected with six links, producing 
an unconstrained Hilbert space of dimension $2^6$ before the Gauss law constraints are 
imposed. The GL sectors related by shift symmetry are $[(0,0)~{\rm and}~(1,-1)]$, 
$[(2,-2)~{\rm and}~(-1,1)]$, $[(3,-3)~{\rm and}~(-2,2)]$.
For the GL sector $(0,0)$, putting $G=0$ into the above Gauss law constraint equations fixes 
the required divergence on even (odd) sites. Counting all six link electric field 
configurations consistent with these conditions gives $20 \ (15)$ states for $ n = 0 \ (n=1)$ 
on even sites and $15 \ (20)$ on odd sites, yielding in total $35 \ (35)$ solution for even (odd) 
sites for this sector. Similarly, the sector $(1,-1)$ also has $35 \ (35)$ solutions for even
(odd) sites. For the sector $(2,-2)$ we obtain $21 \ (21) $ and for the sector $(3,-3)$ we
get $7 \ (7)$ solutions. There is another GL sector $(4,-4)$ (and its shifted partner $(-3,3)$) 
where there is a single state of completely frozen fermions which do not hop at all.

 How do we know that we have considered all the GL sectors? This can be argued through a
a general counting formula. Consider a hypercubic lattice in spatial dimension $d$, so each 
site touches $2d$ links. The total number of all possible local states is
\begin{equation}
    N_{\rm total}(d) = 2^{2d} = \sum_{k=-d}^d {2d \choose d+k},
\end{equation}
spread across all possible GL sectors. Each link $S^z_\ell = \pm 1/2$. 
With our GL conventions, for a fixed value of the local GL $g$ and a fixed matter occuptation $n_x$,
the number of local solutions are 
\begin{equation}
    N(d,g,n_x) = {2d \choose d+n_x-g}
\end{equation}
where $\nabla.E = n_x - g$. The total number of solutions in a sector can be obtained by summing 
over $n_x=0,1$. For example, in $d=3$, this gives it gives 35, 21, 7, 1 solutions for GL sector 
$(0,0)$, $(2,-2)$, $(3,-3)$, and $(4,-4)$ respectively, consistent with the local Gauss law constraints.
This clearly also implies that these are all the possible solutions since they add up to $64 = 2^6$,
which are all the possible link orientations for the electric fluxes in this representation.

\subsection{III. Sign problem in different GL sectors} \label{sec:GL0}
\begin{figure*}[t]
 \centering
 \hspace{-0.4cm}
 \includegraphics[width=0.49\linewidth]{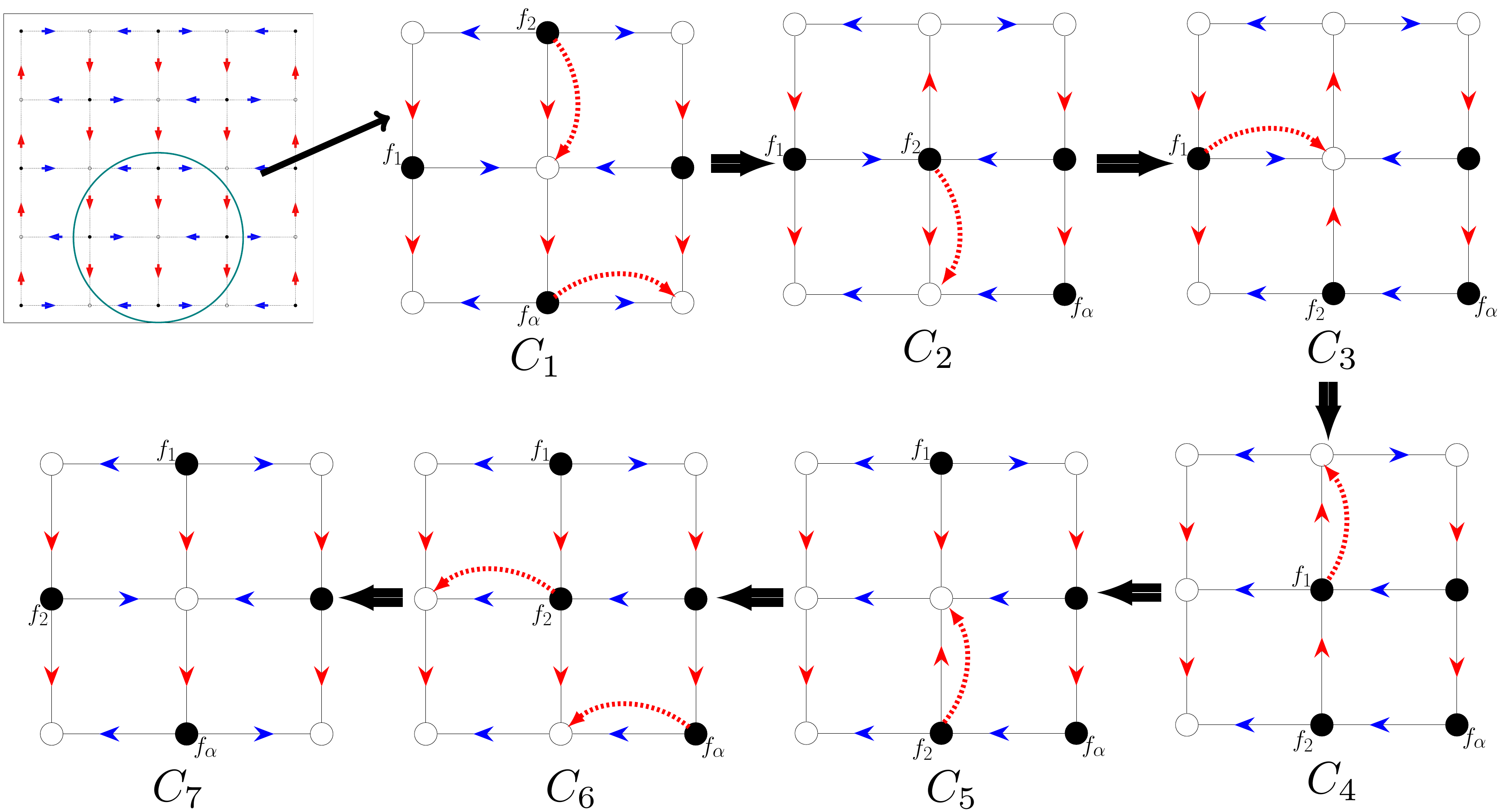}
 \hspace{0.2cm}
 \includegraphics[width=0.49\linewidth]{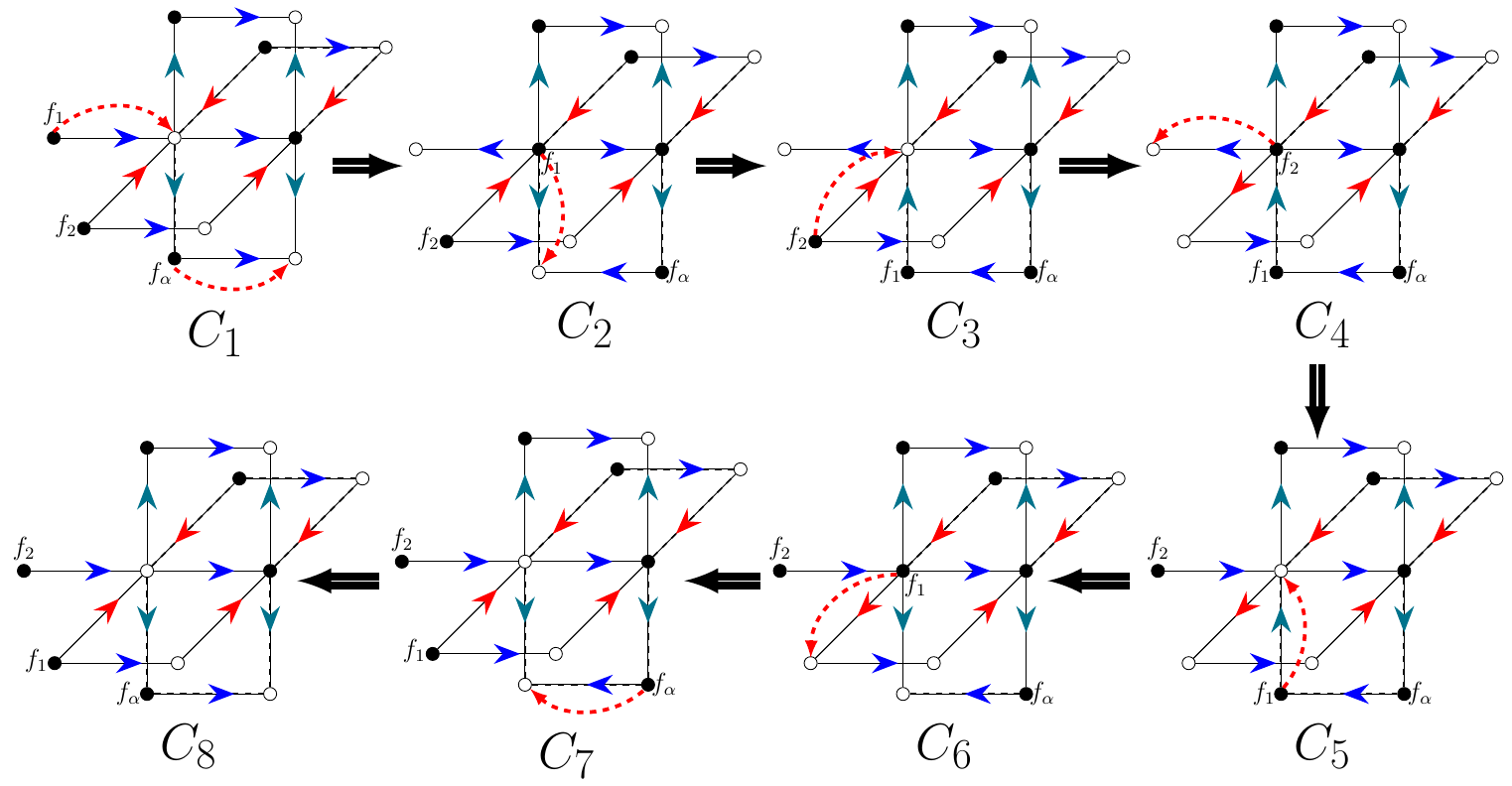}
  \caption{The GL constraints in the sector $(0,0)$ in $d=2$ (left) and $d=3$ (right) 
  are relaxed enough to allow fermions $f_1$ and $f_2$ to exchange positions with each 
  other following giving rise to the fermion sign problem in these sectors.}
  \label{fig:GL0}
\end{figure*}

 In the main text, we discussed a procedure on how to examine whether a particular GL sector
suffers from a sign problem. Using this procedure, we now demonstrate that the physical sector
of interest to particle physics, which is often taken to be $(0,0)$, suffers from a sign problem
in both $d=2,3$. The proof is outlined pictorially in \cref{fig:GL0}, which attempts to switch 
the position of two fermions in a configuration keeping the location of all other fermions fixed.

\subsection{IV. Cluster Algorithm} \label{sec:qmc}
 In this subsection, we discuss the construction of the quantum Monte Carlo algorithm,
which was used to obtain results for larger lattices. The algorithm is constructed 
in terms of worldline configurations of the fermion occupation basis ($n_x$) and the 
electric-flux ($s^3 = E$) basis for the gauge links. In any dimension, the Hamiltonian 
is decomposed into a set of operators, such that operators in each set mutually commute.
In $d=3$, the Hamiltonian is decomposed into six components,
\begin{equation}
    H = H_1 + H_2 + H_3 + H_4 + H_5 + H_6,
\end{equation}
with 
\begin{equation}
    \begin{aligned}
        H_i &= \sum_{\substack{x= (x_1, x_2, x_3) \\ x_i \in \text{even}}} h_{x,i}, &\quad
        H_{i+3} &= \sum_{\substack{x = (x_1, x_2, x_3) \\ x_i \in \text{odd}}} h_{x,i}.
    \end{aligned}
\end{equation}
This Suzuki-Trotter decomposition \cite{Chandrasekharan_2000} allows an expansion of the Boltzmann operator 
$\exp(-\beta H)$ across $2dN_t = \beta$ discrete Euclidean time slices of spacing 
$\epsilon = \frac{\beta}{N_t}$. The partition function then takes the form 
\begin{equation}{\label{eq:Z}}
\begin{aligned}
   Z &= \mathrm{Tr}(e^{-\beta H})\\
     &= \sum_{\{s, n\}}
     \prod_{\tau}
     \bra{ s(x,\tau), n(x,\tau)}  e^{-\epsilon H_1} \\ &\quad \cdots e^{-\epsilon H_6} 
     \ket{ s(x,\tau-1), n(x,\tau-1)},
\end{aligned}
\end{equation}
where $\ket{ s(x,0), n(x,0)}=\ket{ s(x,\beta), n(x,\beta)}$. In this representation, 
the occupied sites define the fermion worldline that are closed 
in Euclidean time, and electric flux variables track the gauge field dynamics. Each Trotter 
slice encodes a local configuration of fermion occupation and link variable, and the 
sequential operation of the transfer matrix, $e^{-\epsilon H_i}$, generates the full 
worldline ensemble. These worldline configurations form the basis for constructing clusters. 
Evaluating the matrix elements of the transfer matrix allows the partition function to be 
expressed in terms of a local action and sign factor, where the contribution from each 
plaquette takes the form 
\begin{align}
  W_{\rm plaq} &= e^{-S[n(x,\tau),n(x+i,\tau),n(x,\tau+1),n(x+i,\tau+1); s^3(x,\tau),s^3(x,\tau+1)]},\\
  Z            &= \prod_{x,\tau} \sum_{n(x,\tau),s^3(x,\tau)} {\rm sign \left[ \{n \}\right] }~ W_{\rm plaq}.
\end{align}

Here, the local action $S[n(x,\tau),n(x+i,\tau),n(x,\tau+1),n(x+i,\tau+1);s^3(x,\tau),s^3(x,\tau+1)]$ 
couples the fermion occupations on neighbour sites at $\tau$ with their neighbours 
in $\tau+1$, as well as the corresponding electric flux variables in these time slices,
while ${\rm sign \left[ \{n \}\right] }$ tracks negative signs while for fermionic worldlines.
\begin{figure}
    \centering
    \includegraphics[scale=0.7]{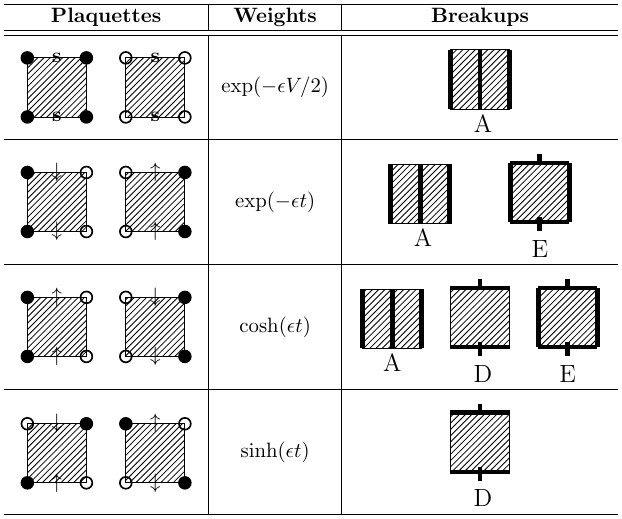}
    \caption{Breakup and corresponding weights for the spin-$\frac{1}{2}$ U(1) gauge links 
    coupled with matter Hamiltonian in \cref{eq:H_U1}.}
    \label{fig:breakup}
\end{figure}
Each active plaquette (shaded in gray) must correspond to one of the configurations listed 
in \cref{fig:breakup}. The weight of plaquette is computed from the local matrix elements 
$\braket{s_b, n_b | e^{-\epsilon H_b}| s'_b, n'_b }$, where $b$ is a nearest-neighbor bond, 
$b = {\{x, x+i}\}$. By introducing appropriate breakups associated with each plaquette, 
the matrix element products from \cref{eq:Z} can be decomposed into clusters, which preserve the allowed 
local configurations while maintaining detailed balance. 

\emph{Breakup} is a technical term indicating that a subset of the six degrees of freedom 
in an active plaquette (four fermions at the two corners, and two gauge fluxes in the 
joining bonds) are connected together in a manner that satisfies detailed balance and can 
be updated simultaneously. Breakups can be of different types, which connect different subsets. 
For example, in order to satisfy detailed balance, the A, D and E breakups shown in 
\cref{fig:breakup}, need to be applied with a probability given by \cref{eq:detailedbalance} 
for $V\geq 2t$. Once breakups are applied at all active plaquettes in a configuration, 
one can define a cluster as the set of all fermions and gauge links which are connected
together. Meron clusters, or merons, are those which on flipping changes the sign of weight of the 
configurations. 

In this formulation, the link variables also contribute, either as additional lines in the 
A breakups or as binding extensions in the D and E breakups, representing a key generalization 
of the original meron cluster algorithm to gauge theories.
\begin{equation}\label{eq:detailedbalance}
\begin{aligned}
    W_A &= \exp(-\epsilon V/2) \\
    W_D &= \sinh(\epsilon t) \\
    W_{E} &= \exp(-\epsilon t) - \exp(-\epsilon V/2). 
\end{aligned}
\end{equation}

 Flipping a cluster is defined as an operation that exchanges the occupied (up) and empty (down) 
sites (links). If a cluster is flipped, the magnitude of the weights of configuration remain 
the same. But if the cluster flip changes the sign of the configuration, it is a meron.
Rules for identifying a cluster as a meron are given in the next section.
With a certain choice of breakups it is possible to factorize the sign factor of a configuration 
$C$ into a product of the signs of each cluster, $\text{Sign}[C] = \prod_i^{N_c}\text{Sign}[C_i]$, 
where $C$ decomposes into $N_c$ clusters. By suitably flipping the clusters, one can reach the 
reference configuration with $\text{Sign} = 1$. The QMC update is as follows:
\begin{enumerate}
    \item Start from the reference configuration that contains only the fermion worldlines. 
    Choose an active plaquette randomly.
    \item If a random plaquette can switch to a different breakup, change it with a probability 
    based on its breakup weight.
    \item After a breakup is modified, re-examine the resulting configuration. If the modification 
    leads to merons in the configuration, then restore the breakup to its previous state and return 
    to step 2.
    \item For each cluster, flip all fermion occupations and flux variables with probability $1/2$.
\end{enumerate}

\subsection{V. Meron rule} \label{sec:meron}
\begin{figure}
    \centering
    \includegraphics[scale=0.25]{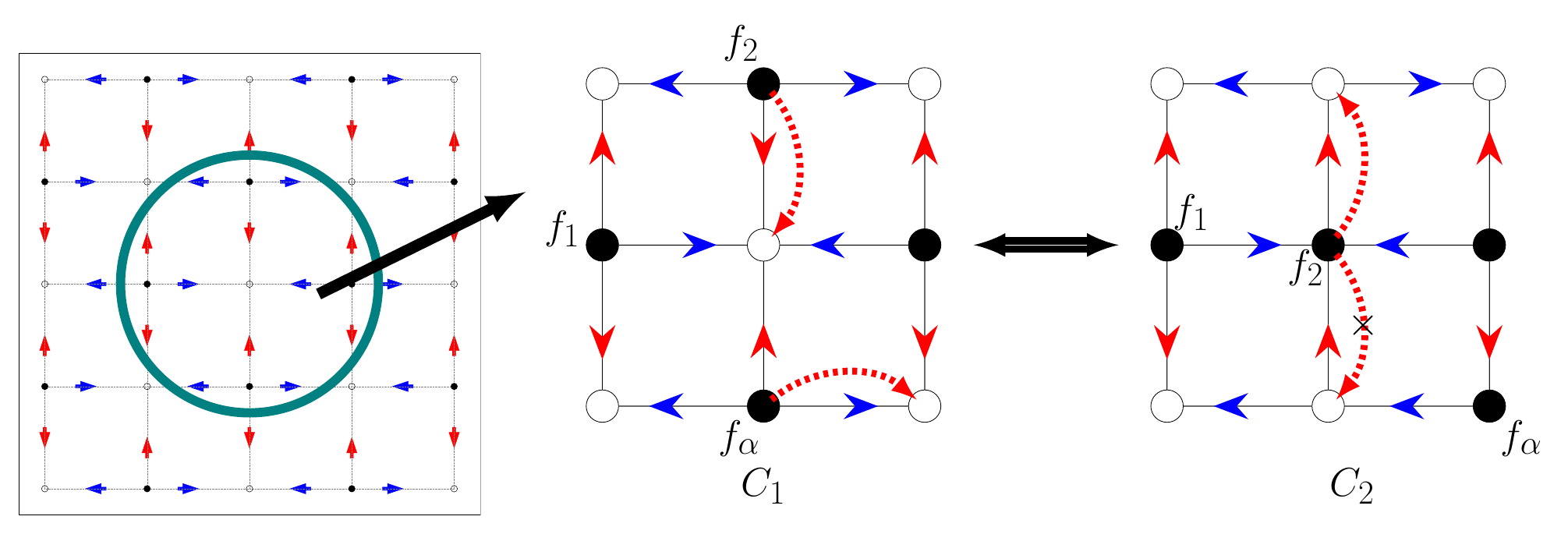}
    \caption{In the Ground-state configuration of the Gauss law sector 
    $G(2,-2)$ in 2-d, the orientation of gauge links restricts the hopping 
    of the fermions $f_1$ and $f_2$ to at most one lattice spacing, which 
    implies that position of fermions cannot be exchanged by the action of Hamiltonian.}
    \label{fig:2d_G2_sign}
\end{figure}

For spinless non-relativistic fermions, the rule \cite{Chandrasekharan1999} of identifying a meron cluster is, 
\begin{equation}
\begin{aligned}
n_w + \frac{n_h}{2} &= \text{even}, & \quad & \text{PBC} \\
n_w + \frac{n_h}{2} &= \text{odd}, & \quad & \text{APBC}.
\label{eq:meronrule1}
\end{aligned}
\end{equation} 
where, $n_w$ is the temporal winding number of the cluster, $n_h$ is the number of hops that 
the cluster makes to neighboring sites while the cluster encounters horizontal bonds 
(D-breakup or E-breakup). PBC and APBC are periodic and anti-periodic boundary conditions,
respectively. When coupled to U(1) gauge fields, the presence of gauge fields allows multiple 
fermion loops to become bound together through the gauge link variables. In that case, the 
following meron definition rules \cite{Banerjee:2023cvs} work for PBC, 
\begin{equation}
\begin{aligned}
n_w + \frac{n_h}{2} &= \text{even}, & \quad & \text{odd \# of loops} \\
n_w + \frac{n_h}{2} &= \text{odd}, & \quad & \text{even \# of loops}.
\label{eq:meronrule2}
\end{aligned}
\end{equation} 
If a cluster contains one fermion loop, the number of loops is odd and the 
first equation of \cref{eq:meronrule2} applies. The condition then reduces to 
\cref{eq:meronrule1}, reproducing the meron rule for the purely fermionic case.
In $d=2,3$, the ground state has no merons, as illustrated in main 
text and also in \cref{fig:2d_G2_sign}. This indicates that the $G(2,-2)$ sector in 2-d 
is free from the fermionic sign problem. 

\subsection{VI. Check with ED}\label{sec:data}
 We have implemented the meron cluster algorithm in both $d=2$ and $d=3$ and benchmarked 
the quantum Monte-Carlo code with the exact diagonalization (ED) result in both cases. 
Naturally, the ED works for small lattice sizes, since the number of Fock states 
increase exponentially. For lattice sizes $2\times8$, $4\times4$, and $2\times2\times2$,
it is easy to do the full ED. For larger systems, we used a Krylov-based framework. We 
constructed the Fock states within the Krylov subspace and performed diagonalization 
using the Lanczos algorithm \cite{lanczos1950} for the lowest 100 states, and then monitored the convergence
of the eigenvalues as a function of the number of Krylov subspaces. The results displayed
in our plots for larger lattices only show results which have fully converged. The
convergence is at least $10^{-4}$ for lattices smaller than the $2 \times 10$ lattice, while an
accuracy of $10^{-3}$ is reached for the larger lattices $2 \times 12$, $4 \times 6$. Of
course, for smaller lattice where full ED could be used, we have machine-precision results.

 \begin{table}[H]
        \centering
        \resizebox{0.5\textwidth}{!}{
        \begin{tabular}{c | c c | c c }
        \hline
        ${L_x} \times L_y $ & \multicolumn{2}{c|}{(0,0)} & \multicolumn{2}{c}{(2, -2)} \\
         &   k-iter & \# of states & k-iter & \# of states  \\
        \hline
        \hline
         & 4 & 17456 & & \\
         & 6 & 238467 & & \\
         & 7 & 686977 & & \\
         & 8 & 1744187 & & \\
         & 9 & 3984067 & & \\
         $2 \times 10 $ & 10 & 8298344 & 10 & 891097\\
         & 11 & 15935644 & & \\
         & 12 & 28439159 & & \\
         & 13 & 47474769 & & \\
         & 14 & 74495344 & & \\
        \hline 
        & 3 & 5935 & & \\
        & 4 & 39670 & & \\
        & 5 & 199198 & & \\
        & 6 & 807368 &  &\\
        $2\times12$ & 7 & 2773220 & 12& 13800964 \\
        & 8 & 8320787 & & \\
        & 9 & 22280495 & & \\
        & 10 & 54105383 & &\\
        & 11 & 120566831 & & \\
        \hline
        & 4 & 35254 & & \\
        & 5 & 165754 & & \\
        & 6 & 630644 & & \\
        & 7 & 2048564 & & \\
        & 8 & 5853611 & & \\
       $4\times6$ & 9 & 15026683 & 12 & 8205424 \\
        & 10 & 35186815 & & \\
        & 11 & 75957235 & & \\
        & 12 & 152451955 & & \\
        & 13 & 286253539 & & \\
        \hline
        & 3 & 21016 & 3 & 47869 \\
        & 4 & 220825 & 4 & 637027\\
        $6\times6$ & 5 & 1723735 & 5 & 6023779 \\
        & 6 & 10572547 & 6 & 41850295 \\
        & 7 & 53331511 & 7 & 218048551 \\
        & & & 8 & 863252872 \\
        \hline
        \end{tabular}
        }
        \caption{The number of allowed states in the $G(e,o)$ sectors in 2D under krylov space expansion.}
        \label{tab:krylovGsector2d}
    \end{table}   

Let us give an estimate of the matrix sizes involved in the procedure. The unconstrained 
Hamiltonian with spin-$\tfrac{1}{2}$ gauge links coupled to matter on a lattice of volume 
$\mathcal{V} = L_x \times L_y$ in 2D and $\mathcal{V} = L_x \times L_y \times L_z$ in 3D 
contains $2^{3\times\mathcal{V}}$ and $2^{4\times\mathcal{V}}$ possible configurations 
respectively. Imposing the Gauss law constraints in order to select sectors defined by
$(G_e,G_o)$, significantly reduces the Hilbert space dimension. In our simulations, we 
focus on the Gauss law sectors $(0,0)$ and $(2,-2)$ in $d=2$, and $(0,0)$, $(2,-2)$, and 
$(3,-3)$ in $d=3$. The corresponding accessible Hilbert space dimensions obtained via this 
method are listed in \cref{tab:krylovGsector2d} and \cref{tab:3dGsector}.

\begin{table}[H]
        \centering
        \resizebox{0.5\textwidth}{!}{
        \begin{tabular}{c | c | c c | c }
        \hline
        ${L_x} \times L_y \times L_z$ & (0,0) & \multicolumn{2}{c|}{(2, -2)} & (3, -3) \\
         &  & k-iter & \# of states &  \\
        \hline
        \hline
        $2 \times 2 \times 2$ & 303721  & & 42393   & 689\\
        \hline 
        &  &  10 & 22009434 &  \\ 
        $2 \times 2 \times 4$  & & 12 & 85532652 & 400481 \\
        & & 15 & 377993792 & \\
        \hline
        \end{tabular}
        }
        \caption{The number of allowed states in the $G(e,o)$ sectors in 3D.}
        \label{tab:3dGsector}
    \end{table}
    
\begin{figure}[H]
    \centering
    \includegraphics[width=\linewidth]{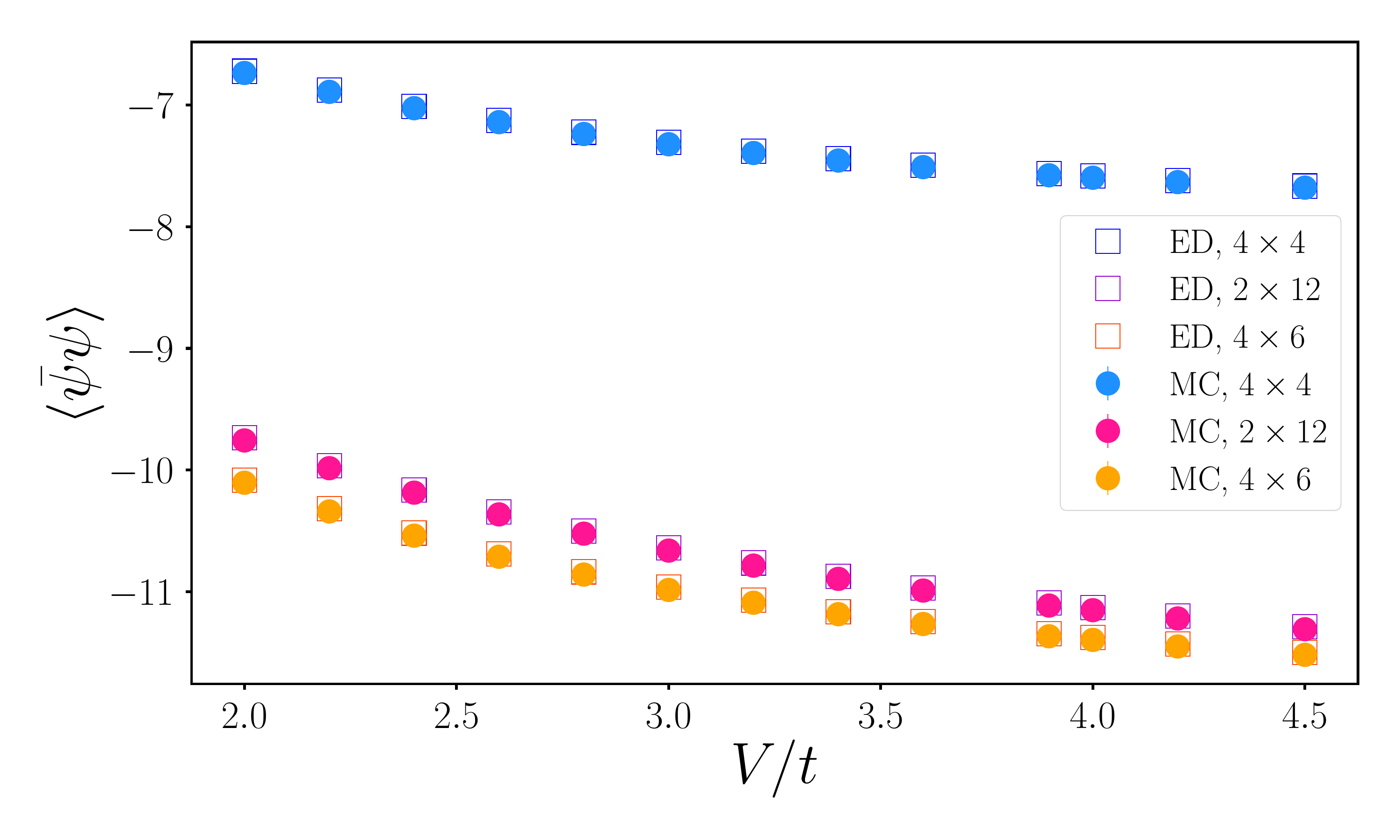}
    \caption{Comparison between meron-cluster (MC) and exact diagonalization (ED) 
    results for the observable chiral condensate $\langle \psi^\dagger \psi \rangle$ 
    with interacting potential $V/t$.}
    \label{fig:chiral condensate}
\end{figure}

We have checked the following observables against ED calculations with results computed
using the Monte Carlo algorithm. The results are summarized in \cref{tab:checkED_2D}, 
\cref{tab:checkED_3D} for $V=2t$ in $d=2$ and $d=3$ respectively, and in 
\cref{fig:chiral condensate} with $V/t$ in $d=2$. All observables are evaluated in the 
ground state Gauss law sector $G(e,o) = (2,-2)$ in $d=2$ and $G(e,o) = (3,-3)$ in $d=3$ 
respectively. 

\begin{equation}\label{eq:observable}
    \begin{aligned}
      \psi^\dagger \psi &= \frac{1}{L_t}  \sum_{x,t} (-1)^x n_x \\
      \epsilon_\mu &= \frac{1}{L_t}  \sum_{x,t} (-1)^x S^3_{x,x+\hat{\mu}}, \quad \mu, \nu =1, 2,\dots, D\\
      \mathcal{O} &=  \frac{1}{L_t} \sum_{x,t} (U_\Box + U_\Box^\dagger) , \quad  U_\Box = U_{x,\mu}U_{x+\hat \mu,\nu} U^\dagger_{x+\hat \nu,\mu} U^\dagger_{x, \nu} \\
      C(r) &= \frac{1}{\cal V}\sum_{x,t} (-1)^r (n_x - \tfrac12)(n_{x+r} - \tfrac12). 
    \end{aligned}
\end{equation}
\begin{table}[H]
    \centering
    \resizebox{0.5\textwidth}{!}{
    \begin{tabular}{c|c|c|c|c|c}
    \hline
    ${L_x} \times L_y$ & & $\langle \psi^\dagger \psi \rangle$ & $ \langle \epsilon_x \rangle $ & $ \langle \epsilon_y \rangle $ & $ \langle \mathcal{O} \rangle$ \\
    \hline
    \hline
    $2\times 6$ & ED &  -4.8682  & -5.6576 & -5.7765 & 0.02395\\
    & MC & -4.8681(5) & -5.6575(3) & -5.7766(1) & 0.02396(5)\\
    \hline
    $2\times8$ & ED &  -6.4909 & -7.5435 & -7.7020 & 0.03193 \\
    & MC & -6.4912(6) & -7.5433(3) &-7.7023(2) & 0.03199(9)\\
    \hline
    $4\times4$ & ED & -6.72205 & -7.6805 & -7.6805 & 0.01816\\
    & MC & -6.7221(12) & -7.6804(4) & -7.6807(5) & 0.01812(8) \\
    \hline
    $2\times12$ & ED & -9.736 & -11.315 & -11.553 & 0.0479 \\
    & MC & -9.737(2) & -11.316(1) & -11.552(1) & 0.0480(1) \\
    \hline
    $4\times6$ & ED & -10.085 & -11.5206 & -11.5217 & 0.02716
    \\
    & MC & -10.084(1) & -11.5203(3) & -11.5219(2) & 0.02723(5)\\
    \hline
    \end{tabular}
    }
    \caption{Comparison of Meron Cluster(MC) values with the ED values 
    for the observables  $\langle \psi^\dagger \psi \rangle$,  
    $ \langle \epsilon_x \rangle $,  $ \langle \epsilon_y \rangle $ and 
    $ \langle \mathcal{O} \rangle$ for $d=2$, and defined in \cref{eq:observable}.}
    \label{tab:checkED_2D}
\end{table}

\begin{table}
    \centering
    \resizebox{0.5\textwidth}{!}{
    \begin{tabular}{c|c|c|c|c}
    \hline
    ${L_x} \times L_y \times Lz$ & & $\langle \psi^\dagger \psi \rangle$ & $ \langle \epsilon_x \rangle $ & $ \langle \epsilon_y \rangle $ \\
    \hline
    \hline
    $2\times2\times2$ & ED & -3.529 & -3.922 & -3.922 \\
    & MC & -3.538(6) & -3.924(2) &  -3.923(2) \\
    \hline
    $2\times2\times4$ & ED & -7.109 & -7.841  & -7.841 \\
    & MC & -7.125(11) & -7.841(5) & -7.843(3)\\
    \hline
    \end{tabular}
    }
    \caption{Comparison between meron-cluster (MC) and exact diagonalization (ED) 
    results for the observables $\langle \psi^\dagger \psi \rangle$,  
    $ \langle \epsilon_x \rangle $ and  $ \langle \epsilon_y \rangle $ for $d=3$
    and defined in \cref{eq:observable}.}
    \label{tab:checkED_3D}
\end{table}

\subsection{VII. Finite size scaling in the phase diagram} \label{sec:FSS}
 In the main text, we provided a schematic outline of the phase diagram of the gauge theory 
with fermionic matter in both the GL sectors $(2,-2)$ and $(0,0)$ in $d=2$. We explain
the physics of the different phases in some more detail in this subsection, by looking
at the scaling of the mass gap in different regimes, 
and argue for a possible candidate spin-liquid phase. 
We also provide results from bigger lattices using MC simulations to argue that the 
thermodynamic limit is reached very fast.

\begin{figure}[t]
    \centering
    \includegraphics[width=\linewidth]{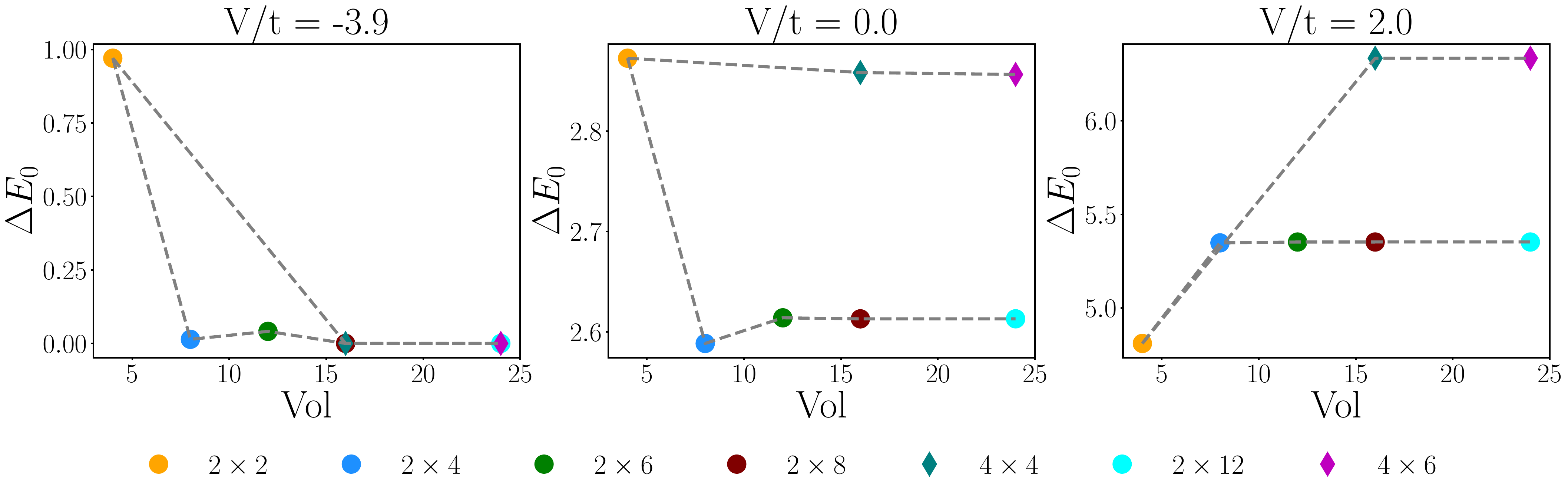}
    \caption{Finite volume scaling behaviour of the gap to the first excited state 
    for fermionic matter in $G(2,-2)$ sector for three different values of $V/t$ 
    corresponding to the three different phases described in the main text.}
    \label{fig:spectra_G2}
\end{figure}

 Let us consider the GL sector $(2,-2)$ first. Based on our result from examining the nature
of the sign problem, we know that since the fermions cannot hop more than one lattice spacing
at a time, we expect the excitations to be gapped. Fig \ref{fig:spectra_G2} shows the behaviour 
of the finite volume mass gap with volume in three different regimes $V/t = -3.9, 0.0, 2.0$. 
For $V/t = -3.9$, the mass gap vanishes, since this corresponds to a regime where the discrete 
rotation symmetry breaks spontaneously. Large negative values of $V/t$ imply that Fock states 
where nearest neighbour states are both occupied or both empty form the ground state. However,
all such states have domain walls separating regions which are completely filled with regions
that are empty in order to maintain the half-filling. This simple considerations imply the 
existence of several degenerate ground state energies, and thus the gap vanishes, as seen in
the panel showing the gap in $V/t=-3.9$. 

 Next, we consider the phase that is stabilized for $V/t \sim 0$. This is the region where
the fermions can maximally hop, but never far. Consequently, the gap is finite and $O(1)$,
and shows little finite volume effect. Interestingly, ladder systems with $L_y=2$ saturate 
to a slightly lower gap as compared to those with $L_y=4$. It is likely that the phase is
a massive $Z_2$ gauge theory. For larger positive $V/t$, there is a unique charge density
wave with staggered occupation which forms the ground state in the sector $(2,-2)$, and the
orientation of the electric fluxes is set through the Gauss Law. The gap then corresponds
to creating a single \emph{dipole} defect by switching the occupation of an even-odd pair 
of sites and flipping the link between them, which is the action of the kinetic term. 
For a square lattice, this creates six \emph{bad} bonds at which the energy due to the 
density-density interaction increases from $-\frac{1}{4}$ to $\frac{1}{4}$, so 
$6 \times\frac{1}{2} = 3 $. For the same six bonds, the designer
term does not contribute any more compared to the ground state, while there is no
additional change from the bond which was just flipped (since both the occupations
and the electric flux orientation changes). This implies that we expect an excitation
above the ground state with $\Delta E \approx 6 V/t$ in the strong coupling limit of 
$V/t \rightarrow \infty$. The observed value of slightly more than 6 is in the 
expected ballpark (note that the density-density term comes with a $V-t$ coefficient). 
This calculation in the strong coupling limit suggests that such dipole defects are
energetically unfavourable in this limit. Moreover, note that this calculation 
also suggests the phase to have massive quasiparticles. For the opposite sign of 
the designer term, creation of such dipole defects could be made favourable, thus 
triggering an instability of the ground state leading to a phase transition. 

\begin{figure}[t]
    \centering
    \includegraphics[width=\linewidth]{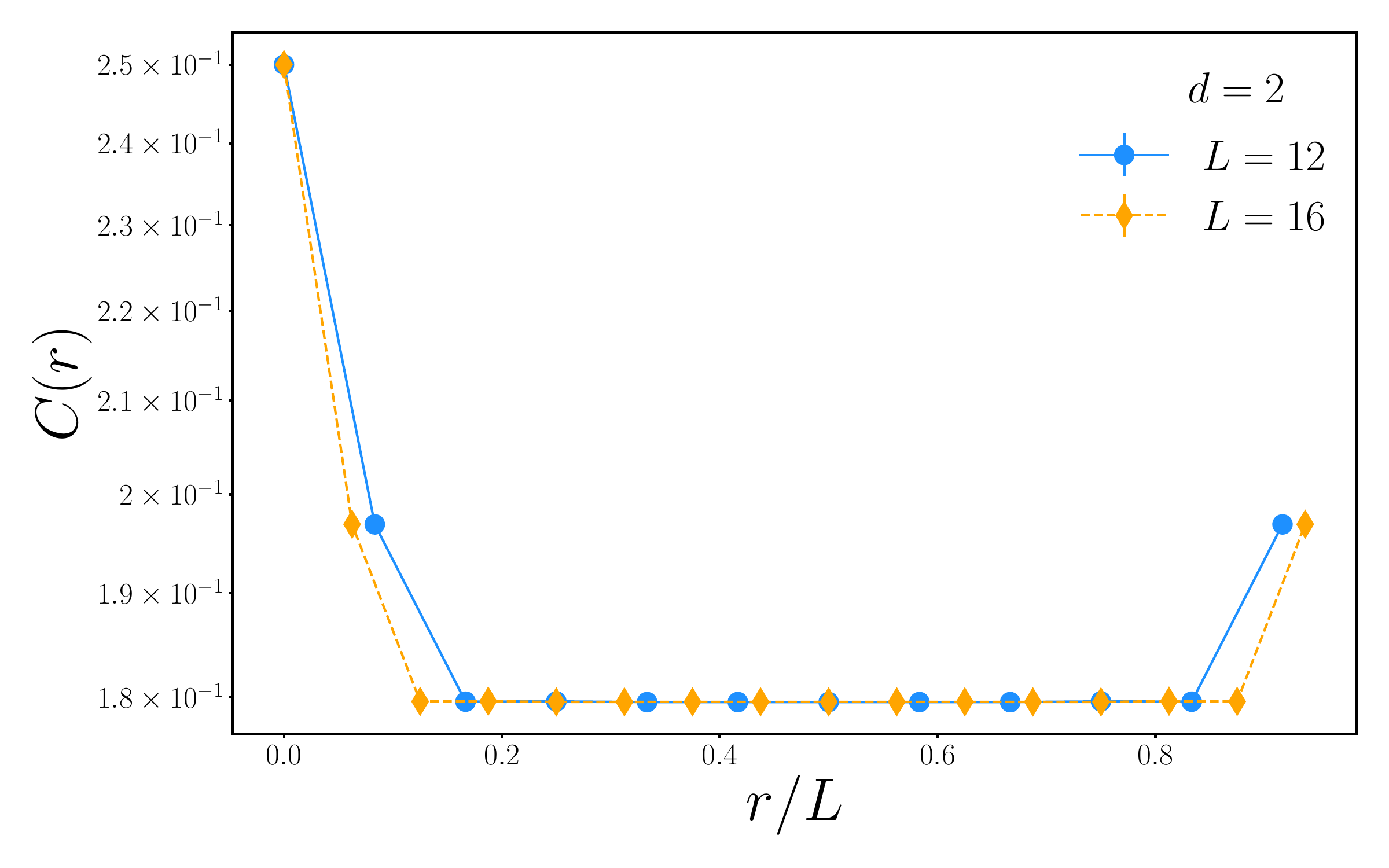}
    \includegraphics[width=\linewidth]{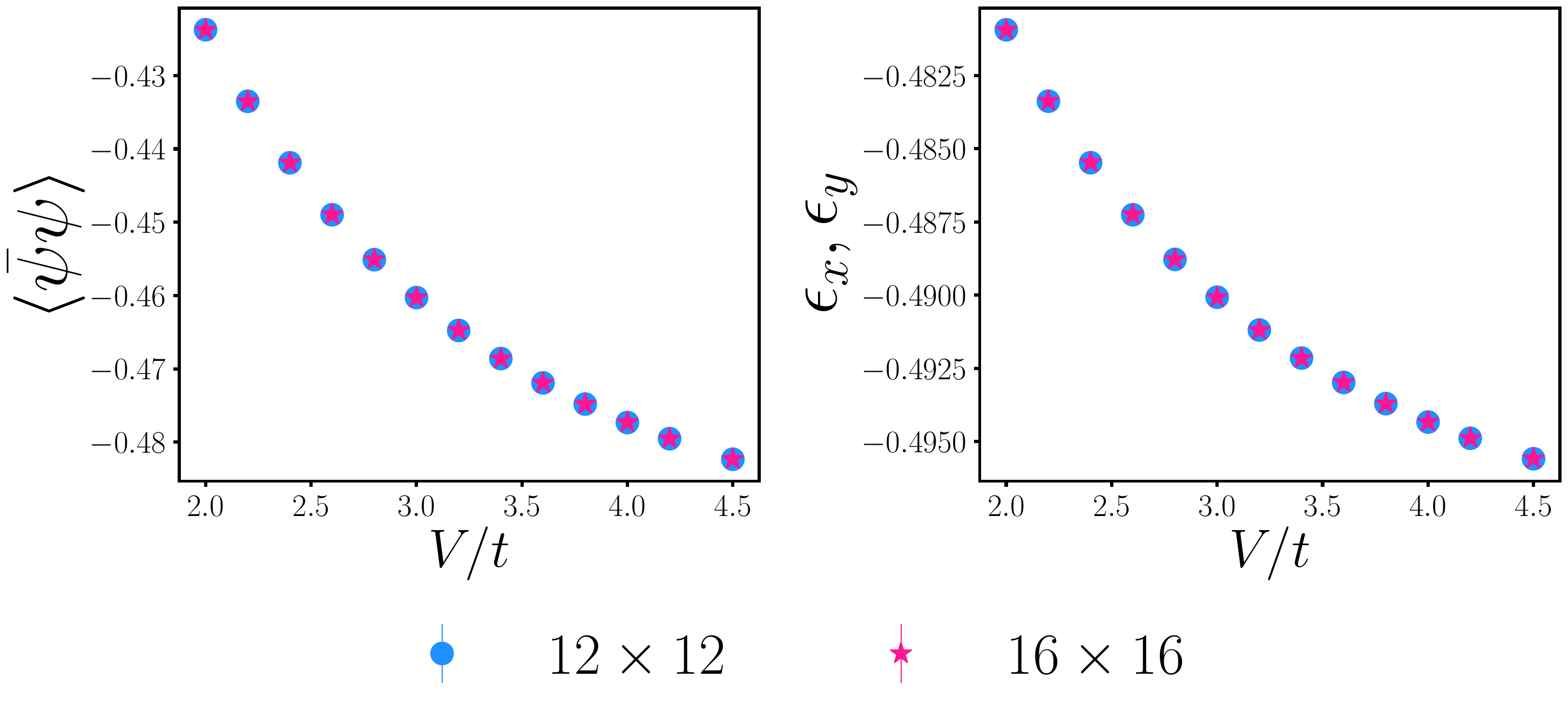}
    \caption{
    (Top) Charge-charge correlation function, (bottom left) 
    $\psi^\dagger \psi$ and  (bottom right) $\epsilon_{x,y}$ from QMC on square lattices 
    of size $L=12,16$.}    
    \label{fig:corrfs_G2}
\end{figure}

Results from the cluster algorithm can simulate the region 
$V \geq 2t$, and we show some observables on lattices of size $L = 12,16$. The 
top panel in \cref{fig:corrfs_G2} shows the charge-charge correlation function
at $V=2t$, from where it is clear that the correlation length is about two 
lattice spacings, and does not scale with the system size, making it clear
that this region in phase space has massive excitations. The left figure in the bottom panel 
shows the chiral condensate, $\braket{\psi^\dagger \psi}$ and the right figure
shows the winding numbers, $\braket{\epsilon_{x,y}}$, all computed on the ground 
state in the $(2,-2)$ sector, showing that the thermodynamic limit has been reached. 
The bottom panel show the observables are shown for a range of $V/t \geq 2$,
and can be seen to asymptote to the value at $V/t \rightarrow -\infty$, -0.5. 

\begin{figure}[t]
    \centering
    \includegraphics[width=\linewidth]{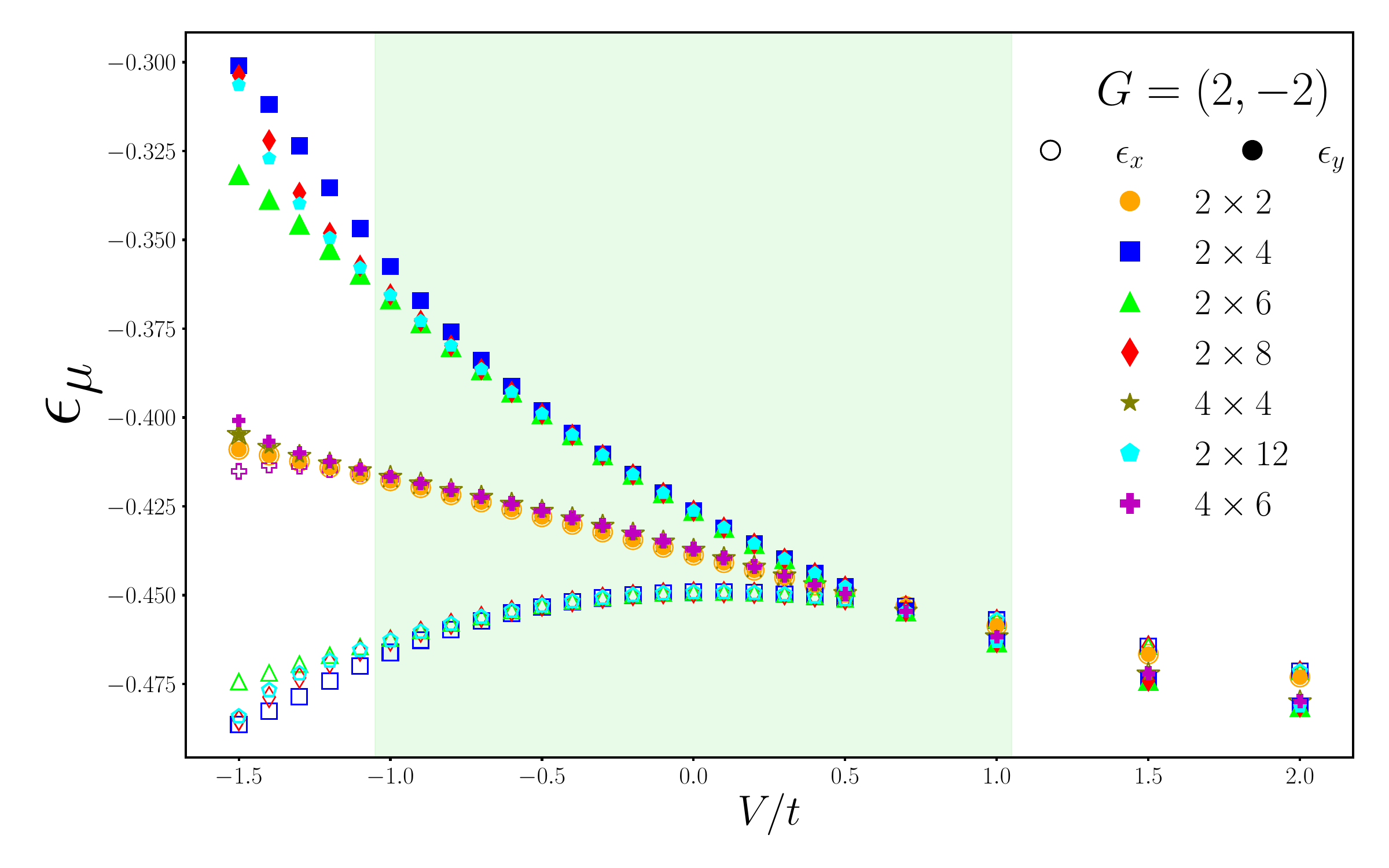}
    \caption{
    FSS of different lattice sizes, with both ladder and square
    geometry shows that the winding numbers $\epsilon_x$ and $\epsilon_y$ become
    comparable suggesting that the rotational symmetry is restored. This is the
    region in the parameter space where a liquid could exist.}
    \label{fig:sliq}
 \end{figure}

 Although the cluster algorithm cannot be used for $V/t < 2$, the fast approach to
 the thermodynamic limit makes the ED results on small lattices trustworthy. We already
 provided a sketch of the phase diagram for the sector $(2,-2)$ in \cref{fig:phaseDiag}
 of the main text. The most interesting part is in the region $-1 \lesssim V/t \lesssim 1 $,
 where we hypothesize the existance of a massive spin liquid. A detailed FSS on the available 
 lattices shown in \cref{fig:sliq} demonstrates that the rotational symmetry has been restored, 
 but the CDW order does not yet kick in. This is a plausible candidate for a liquid 
 phase with massive fermionic excitations. 
 
\begin{figure}[t]
    \centering
    \includegraphics[width=\linewidth]{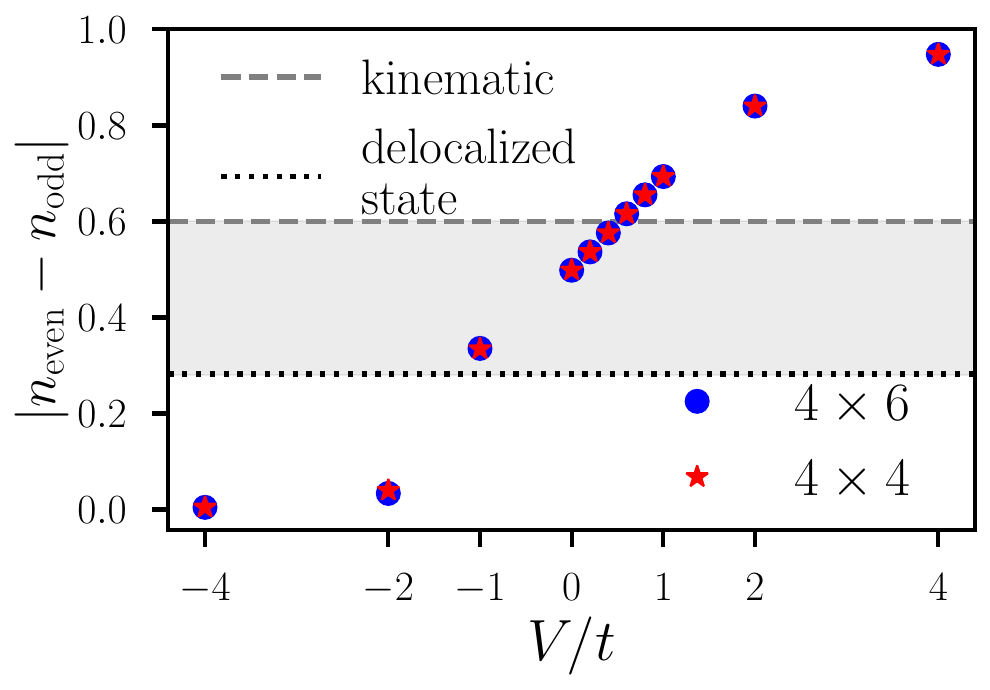}
    \caption{
    The fermion occupation number imbalance computed as a function of different couplings
    on $4 \times 4$ and $4 \times 6$ lattices. The shaded region is bounded by two heuristic
    reference estimates for a maximally disordered state with the constrained $(2,-2)$ Hilbert
    space: an equal local weighting of the allowed Gauss law vertices and an equal-amplitude
    superposition over the complete finite-volume Fock basis.     
    }
    \label{fig:imbalance}
 \end{figure}

\begin{figure}[t]
    \centering
    \includegraphics[width=\linewidth]{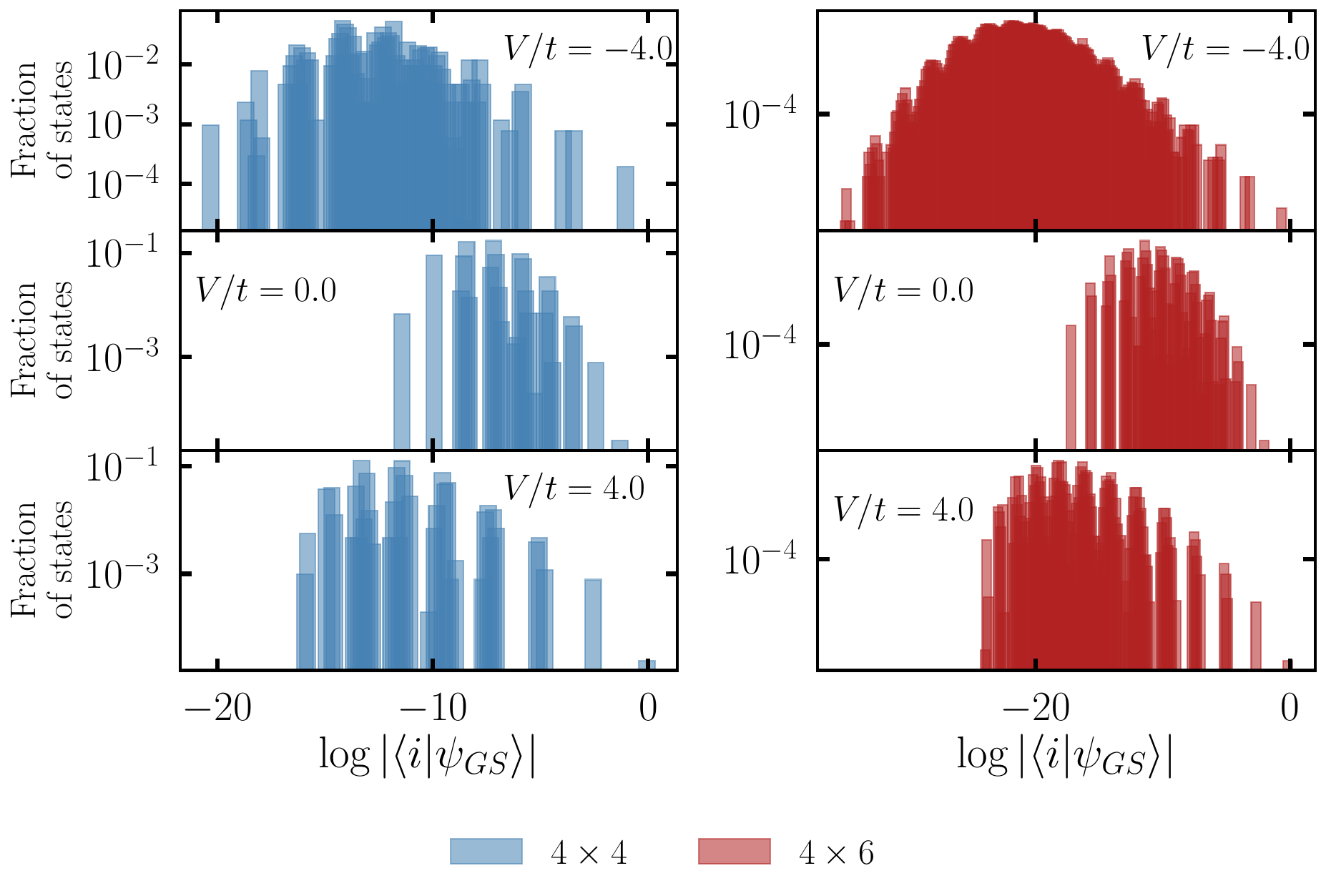}
    \includegraphics[width=\linewidth]{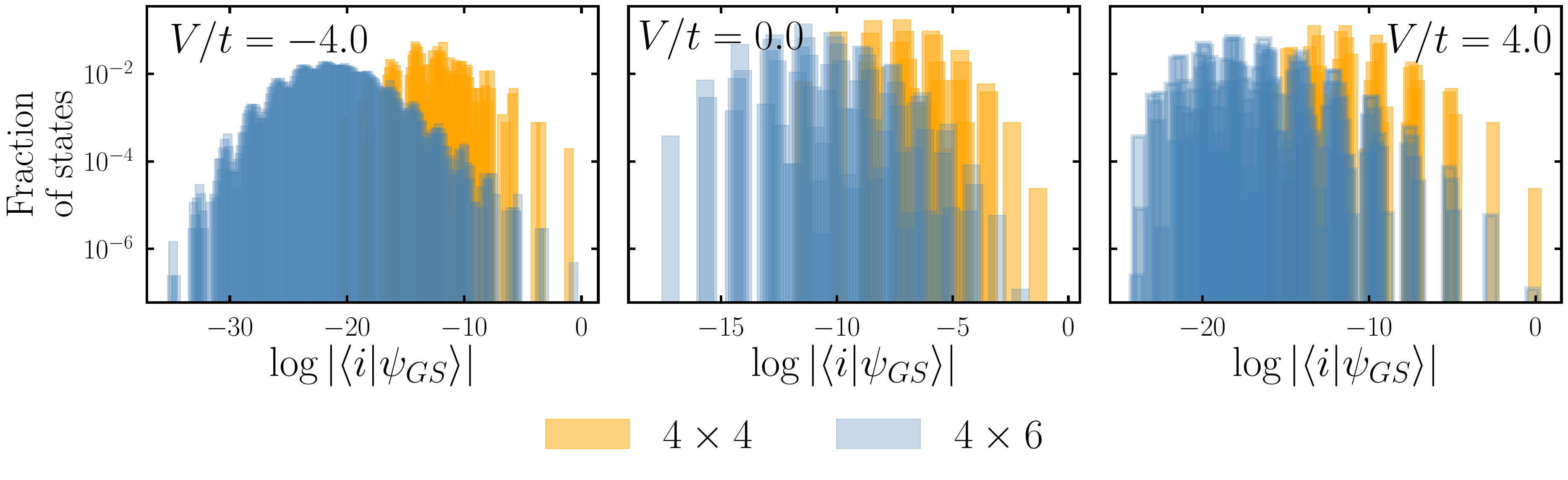}
    \caption{
    (Top) Histogram showing the fraction of the ground state wavefunction in 
    different Fock states. Ordered phases (top, $V/t=-4$ and bottom, $V/t=4$) have 
    a distinct bin close to the origin showing that a few states carry a significant 
    part of the probablity, while the disordered phase (middle, $V/t=0$) has no such
    bin. 
    (Bottom) A finite size scaling of the histograms between $4\times4$ and $6\times4$
    showing that the latter becomes tigher, particularly for $V/t = 0$; and the 
    bins move away from zero, showing that no particular Fock states contribute to
    the ground state wavefunction, unlike for the other two $V/t$ values.
    }
    
    \label{fig:histGS}
 \end{figure}

It is plausible to suspect the existence of a spin-liquid phase in the 
 region around $V/t \sim 0$, in between the two ordered phases, where the fermions
 can be as mobile as allowed by the Gauss Law constraints. While fig.~\ref{fig:sliq}
 indicates the restoration of the broken rotation symmetries for $V/t \sim 0$, 
 figs.~\ref{fig:imbalance} and \ref{fig:histGS} provide further evidence. 
 In fig.~\ref{fig:imbalance}, we consider the imbalance, $I$, which measures the difference 
 in the occupation between even and odd sites. In the CDW phase (large positive $V/t$)
 this asymptotes to 1, while in the phase separated phase (large negative $V/t$) this
 is zero (since the neighbouring sites are either both empty, or both occupied). 
 In fact, it is useful to ask the possible imbalance signatures of a liquid phase. 
 Consider the following two arguments: \\
 \begin{itemize}
   \item Using only the solutions for the Gauss Law at each site in the constrained space, we
   can compute $\overline{n}_e$ and $\overline{n}_o$ in the sector $(2,-2)$, as shown
   in Sec.\hyperref[sec:GLsoln]{II}. For an even site, there are five solutions, four of
   which have $n_e=1$ and one has $n_e=0$, giving $\overline{n}_e=4/5$. For an odd site,
   the solutions are flipped, $n_o=1$ has one solution and $n_o=0$ has 4 solutions,
   and thus $\overline{n}_o=1/5$; and hence $I = |\overline{n}_e - \overline{n}_e| = 0.6$.
   This assumes that the fermions are free to hop about as allowed by Gauss Law and
   are not subject to any ordering.
   \item A second estimate of the imbalance can be obtained by considering a state 
   which is completely delocalized in the Fock space 
   $\ket{\Phi}=\frac{1}{\sqrt{\cal{N}}}\sum_i \ket{i}$,
   where $\cal{N}$ is the total number of states in the GL sector $(2,-2)$. For both
   the $4 \times 4$ and $6 \times 4$ lattice, we obtain 
   $\overline{n}_e \approx 0.64,~\overline{n}_o \approx 0.35$, and hence the imbalance 
   is $\sim 0.3$. 
 \end{itemize}
 These estimates are represented as a band in fig.~\ref{fig:imbalance}; and we clearly
 see that the residual imbalance in the ground state is of the magnitude naturally
 generated by the Gauss law constrained Hilbert space.
 
 Fig.~\ref{fig:histGS} shows a Fock-space diagnostic which was first introduced in 
 \cite{Rammelmueller2022} to study signature of a spin liquid. A picture of the (de)localization 
 of the ground state is obtained by plotting the histogram of the logarithm of the 
 Fock-space amplitudes in the computational basis. For ordered ground states, most 
 of the weight is contained in a few states, i.e., with amplitudes $\sim {\cal O}(1)$,
 and is shown by distinct bins close to zero. Of course, additionally 
 there is a large tail. However, for a liquid phase the bins close to zero disappear
 and there is a single tight curve with no clear peaks. Clear signatures of both
 the localized phases for $V/t \sim 4$ and $V/t \sim -4$ (the bins close to zero 
 indicate the presence of a small number of Fock states with anomalously large amplitudes),
 and declocalization of the ground state for $V/t \sim 0$ (consistent with enhanced 
 Fock-space delocalization). The bottom panel shows that the histogram
 is tigher for $6 \times 4$ than $4 \times 4$ for $V/t \sim 0$ showing a clear lack
 of order. Thus, this diagnostic also consistently points to the existence of a 
 liquid phase in $V/t \sim 0$ region.

In the next figure, \cref{fig:spectra_G0}, we show the corresponding finite size 
scaling of the mass gap for the same values $V/t$ as in \cref{fig:spectra_G2}
but for the $(0,0)$ GL sector. As in the case before, for negative values of $V/t$ 
one will encounter domain wall states leading to several degenerate ground states 
and a vanishing of the mass gap. The 
significantly interesting case is the middle panel, for $V/t=0$: the small mass
gap for $L_y=2$ ladders decreases further for the $4 \times 4$ lattice, indicating
that a possibility exists for the system to have a light quasi particle mode in this
region. This is consistent with the observation in \cite{Hashizume:2021qbb}, but
studies on larger lattices are necessary to reach a conclusion. It is essential 
to develop the meron cluster algorithm to simulate this sector of the theory
which has a genuine sign problem. For larger positive values of $V/t$, the scaling of
the energy gap is similar, and we are unable to definitely conclude whether the
system goes into a new phase based on this data alone. However, \cref{fig:GLPhys}
seem to indicate that diagonal observables (such as the chiral condensate and the
staggered flux) saturate. This in turn indicates the existence of a different
phase for $V/t \gg 0$. Since the mass gap is even lower in this regime, it could
be the spontaneous breaking of a discrete symmetry. Note that the notion of 
\emph{dipole excitations} in the $(0,0)$ GL sector is complicated by the fact that
these sectors have flippable plaquettes. The presence of such flippable plaquettes
can easily give rise to a closed flux excitation which makes a naive counting of
bonds difficult.

\begin{figure}[!tbh]
    \centering
    \includegraphics[width=\linewidth]{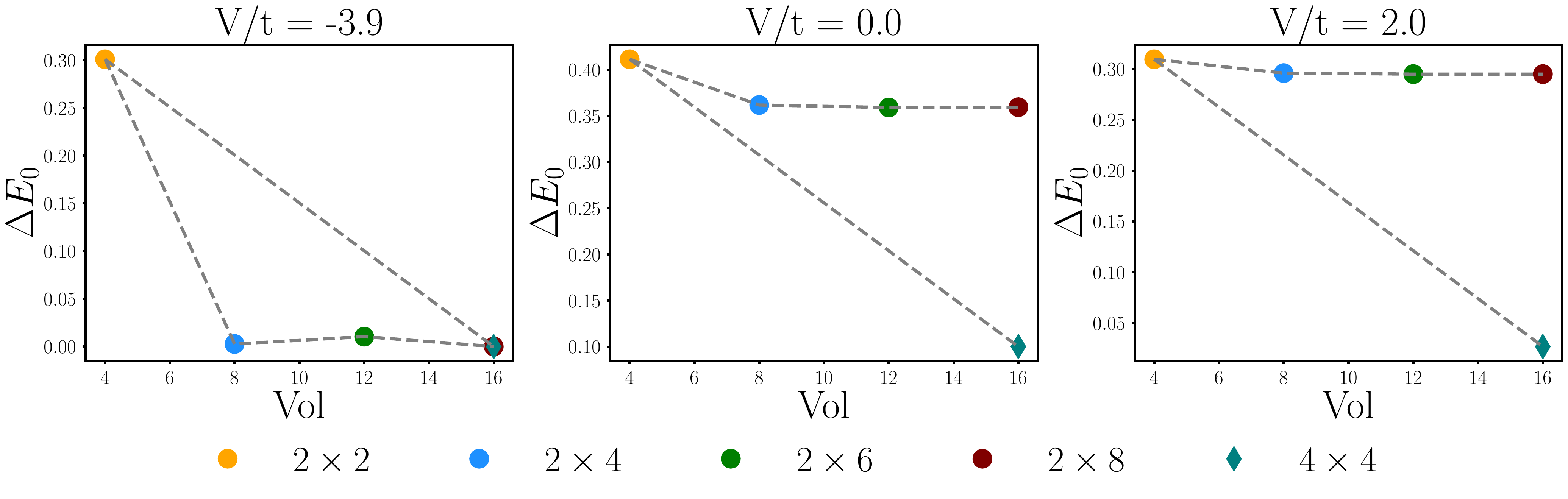}
    \caption{Finite volume scaling behaviour of the gap to the first excited state 
    for the Fermionic case in G(0,0) sector for the same three values of $V/t$ 
    as in the last figure. These again correspond to different phases.}
    \label{fig:spectra_G0}
\end{figure}

\subsection{VIII. Ground-State GL Sectors without designer term}\label{sec:without_designerterm}
\begin{figure}[!tbh]
    \centering
    \includegraphics[width=\linewidth]{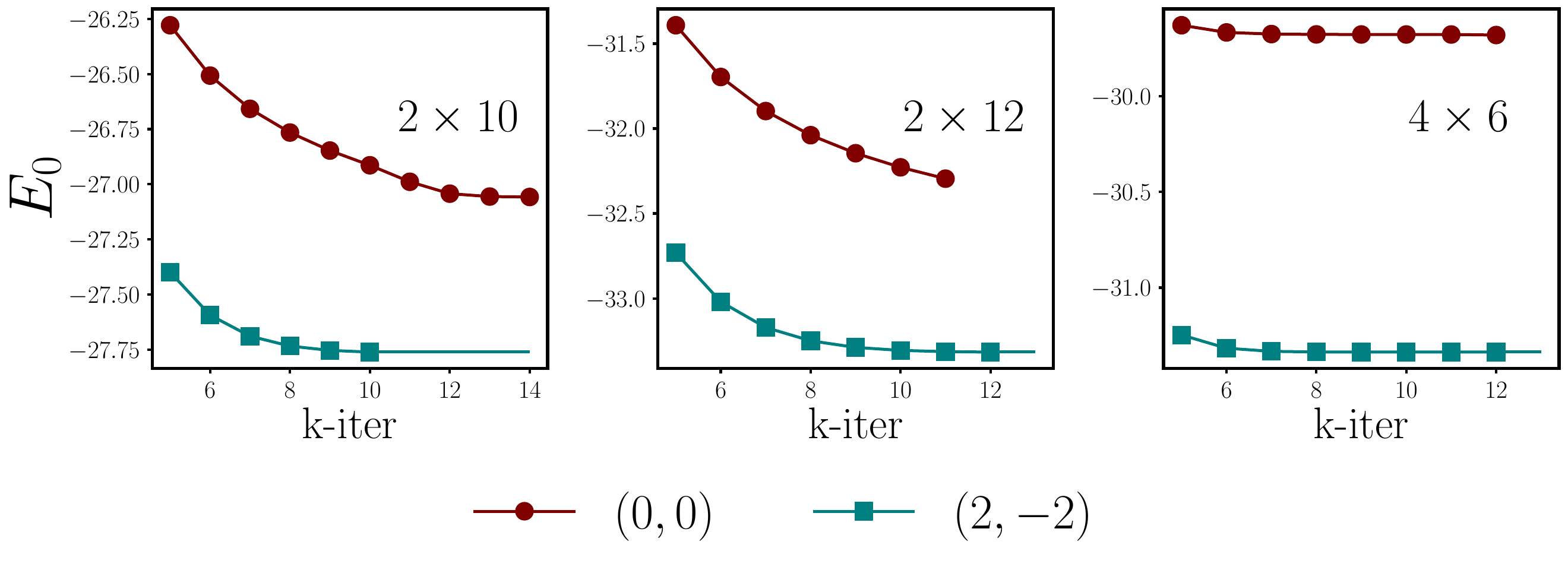}
    \caption{Convergence of Ground state energy for the GL sectors $(0,\ 0)$ 
    and $(2, \ -2)$ with Krylov iteration in the absence of the designer term at $V/t = 2.0$.}
    \label{fig:krylov_Veq2t}
\end{figure}
In the absence of the designer term, we computed the ground-state energy for 
both GL sectors $(0, 0)$ and $(2, -2)$ using a Krylov-based Lanczos method \cite{lanczos1950}. 
In \cref{fig:krylov_Veq2t} we show the convergence of ground state energy $E_0$ as a function 
of Krylov iteration $k$ for lattices $2\times10$, $2\times 12$ and $4\times 6$ at $V/t = 2.0$. 
The $G=(2,-2)$ sector remains at ground-state energy GL sector than $G=(0,0)$, consistent with 
Hamiltonian \cref{eq:H_U1}) with the designer term and also converges faster than the $G=(0,0)$ 
sector owing to its comparatively smaller Hilbert-space dimension across the lattices studied 
\cref{tab:krylovGsector2d}.

\subsection{IX. Relation to quantum dimer and fully packed loop models} \label{sec:mappings}

We now clarify the relation between the spin-$\frac{1}{2}$ $U(1)$ QLM, the square lattice 
quantum dimer model (QDM), and the fully packed loop (FPL) model.  Introducing the site parity
$\epsilon_x=(-1)^{x_1+x_2}$, we write the electric flux in terms of a binary bond variable 
$D_{x,\hat i}=0,1$ as
\begin{equation} \label{eq:sm_flux_dimer_map}
E_{x,\hat i} = \epsilon_x \left( \frac{1}{2}-D_{x,\hat i} \right).
\end{equation}
Let $n_x^d = \sum_i \left( D_{x,\hat i} + D_{x-\hat i,\hat i} \right)$ denote the number of 
occupied bonds touching $x$, then
\begin{equation} \label{eq:sm_divE_dimer}
(\nabla\cdot E)_x = \epsilon_x(2-n_x^d).
\end{equation}
Consequently, in the pure gauge theory, $(\nabla\cdot E)_x=\epsilon_x$ gives the QDM constraint
$n_x^d=1$, whereas $(\nabla\cdot E)_x=0$ gives the FPL constraint $n_x^d=2$.

For dynamical staggered fermions, the full Gauss law operator is
\begin{equation} \label{eq:sm_staggered_GL}
G_x = n_x+\frac{\epsilon_x-1}{2} - (\nabla\cdot E)_x .
\end{equation}
It is useful to introduce the binary matter variable
\begin{equation}
M_x = \epsilon_x \left[n_x-\frac{1-\epsilon_x}{2}\right] =
\begin{cases}
n_x, & \epsilon_x=+1, \\
1-n_x, & \epsilon_x=-1,
\end{cases}
\label{eq:sm_monomer_variable}
\end{equation}
where $M_x\in{0,1}$. Since $n_x+(\epsilon_x-1)/2 = \epsilon_x M_x$, Eqs.~\eqref{eq:sm_divE_dimer} and
\eqref{eq:sm_staggered_GL} give the local master identity 
\begin{equation} \label{eq:sm_master_constraint}
G_x = \epsilon_x \left( n_x^d+M_x-2 \right).
\end{equation}
For a staggered Gauss-law sector $G_x=s\epsilon_x$, the physical Hilbert space therefore obeys
\begin{equation} \label{eq:sm_sector_constraint}
n_x^d+M_x=2+s.
\end{equation}
In particular, the conventional monomer--dimer constraint belongs to the Gauss law sector
\begin{equation}
G_x=-\epsilon_x, \qquad (G_e,G_o)=(-1,1),
\end{equation}
and the fully packed quantum model with monomers belong to 
\begin{equation}
G_x=0, \qquad (G_e,G_o)=(0,0),
\end{equation}

 The shift symmetry discussed in the main text can also be implemented on the monomer and
dimer variables as 
\begin{equation} \label{eq:shift-dimer-matter}
    D'_{x,\mu}=1-D_{x+\hat k,\mu}, \qquad M'_x=1-M_{x+\hat k}.
\end{equation}
Consequently, $n_x^{d\,\prime}=4-n_{x+\hat k}^d$. For example, the QDM-monomer constraint 
is mapped as
\begin{equation}
    n_x^d+M_x=1 \quad\longmapsto\quad n_x^{d\,\prime}+M'_x=4,
\end{equation}
which is precisely the pair $(-1,1)\leftrightarrow(2,-2)$. Likewise, $n_x^d+M_x=2$ is mapped 
to $n_x^{d\,\prime}+M'_x=3$, corresponding to $(0,0)\leftrightarrow(1,-1)$.

 To convince ourselves of this mapping, we have also considered several lattice geometries
and explicitly computed the number of Fock states allowed with the constraints of the 
QDM with monomers ($n_x^d + M_x = 1$) and the FPL model with monomers ($n_x^d + M_x = 2$).
There is an identical mapping with the states in the sector $(-1,1)$ and $(0,0)$ respectively
giving further credence to our proof. The results are shown in Table~\ref{tab:monomer_dimer_counts}.

\vspace{1cm}
\begin{table}[t] 
\centering 
\begin{tabular}{c|r|r} 
\hline\hline 
Lattice    & $n_x^d+M_x=1$ & $n_x^d+M_x=2$ \\ 
           & (-1,1)        & (0,0)         \\
\hline 
$2\times2$ &   17          &    66         \\ 
$2\times4$ &  241          &  3074         \\ 
$2\times6$ & 3716          & 164226        \\ 
$2\times8$ & 57537         & 8946934       \\ 
$4\times4$ & 41025         & 5550554       \\ 
\hline\hline 
\end{tabular} 
\caption{Number of Fock states satisfying the QDM and FPL monomer 
constraints on periodic square lattices.} \label{tab:monomer_dimer_counts} 
\end{table}

Finally, the constraints in both the QDM and the FPL models with monomers imply that two 
monomer degrees of freedom can be annihiliated to form a dimer or vice-versa. This is idential
to the fermion-hopping simultaneous with the flux flip, where possible. 
\begin{equation}
\psi_x^\dagger U_{x,\hat i} \psi_{x+\hat i} +{\rm h.c.} \;\leftrightarrow\;
m_x^\dagger  m_{x+\hat i}^\dagger d_{x,\hat i} + d_{x,\hat i}^\dagger m_{x+\hat i} m_x .
\label{eq:sm_hopping_monomer_dimer}
\end{equation}

\end{document}